\documentclass[journal = cmatex, 12pt, journal=jacsat,manuscript=article]{achemso} 
\usepackage{amsmath} 
\usepackage{tabularx} 
\usepackage{booktabs} 
\usepackage{makecell}
\SectionNumbersOn
\usepackage[utf8]{inputenc}
\usepackage[version=4]{mhchem}

\title{Influence of Electrode Structuring Techniques on the Performance of All-Solid-State Batteries}


\author{Moritz Clausnitzer}
\affiliation[DLR]{German Aerospace Center (DLR), Institute of Engineering Thermodynamic, 70569 Stuttgart, Germany}
\alsoaffiliation[HIU]{Helmholtz Institute Ulm for Electrochemical Energy Storage (HIU),  89081 Ulm, Germany}
\email{moritz.clausnitzer@dlr.de}

\author{Timo Danner}
\affiliation[DLR]{German Aerospace Center (DLR), Institute of Engineering Thermodynamic, 70569 Stuttgart, Germany}
\alsoaffiliation[HIU]{Helmholtz Institute Ulm for Electrochemical Energy Storage (HIU),  89081 Ulm, Germany}

\author{Benedikt Prifling}
\affiliation[UniUlm2]{Ulm University, Institute of Stochastics, 89081 Ulm, Germany}

\author{Matthias Neumann}
\affiliation[UniUlm2]{Ulm University, Institute of Stochastics, 89081 Ulm, Germany}

\author{Volker Schmidt}
\affiliation[UniUlm2]{Ulm University, Institute of Stochastics, 89081 Ulm, Germany}

\author{Arnulf Latz}
\affiliation[DLR]{German Aerospace Center (DLR), Institute of Engineering Thermodynamic, 70569 Stuttgart, Germany}
\alsoaffiliation[HIU]{Helmholtz Institute Ulm for Electrochemical Energy Storage (HIU),  89081 Ulm, Germany}
\alsoaffiliation[UniUlm]{Ulm University, Institute of Electrochemistry, 89081 Ulm, Germany}

\begin{document}

\begin{abstract}

All-solid-state batteries (ASSBs) offer a promising route to safer batteries with superior energy density compared to conventional Li-ion batteries (LIBs). However, the design of the composite cathode and optimization of the underlying microstructure is one of the aspects requiring intensive research. Achieving both high energy and power density remains challenging due to limitations in ionic conductivity and active material loading. Using structure-resolved simulations, we investigate the potential of perforated and layered electrode designs to enhance ASSB performance. Design strategies showing significant performance increase in LIBs are evaluated regarding their application to ASSBs. Composite cathodes with solid electrolyte channels in the structure do not significantly increase cell performance compared to unstructured electrodes. However, the design with a two-layer cathode proves promising. The layered structure effectively balances improved ionic transport due to increased solid electrolyte fraction at the separator side and substantial active material loading through increased active material fraction at the current collector side of the cathode. Our research highlights key challenges in ASSB development and provides a clear direction for future studies in the field.  
\end{abstract}

\section{Introduction}

Electric vehicles (EVs) are an essential component for the decarbonization of the transport sector. The battery chemistry and its corresponding properties substantially influence their environmental impact, cost, and social acceptance. \cite{li2019comprehensive,li2014life,ellingsen2016size}
The weight and efficiency of EVs are directly affected by battery mass. Therefore, lightweight battery cells with high specific energy are of paramount importance for electric vehicles. \cite{diouf2015potential} Additionally, fast charging is a strict requirement for widespread EV adoption, emphasizing the need for cells with high power density. \cite{tu2019extreme, li2020fast} 

Owing to their high energy and power density, Li-ion batteries (LIBs) are the preferred choice for mobile applications. \cite{diouf2015potential} Over the past decade, LIBs have made tremendous progress, achieving energy densities exceeding 250 Wh/kg and allowing fast charging to 80\% state of charge (SOC) in under 30 minutes. \cite{cano2018batteries, placke2017lithium, weiss2021fast} However, additional progress is needed to accelerate the increase in the number of electric vehicles.

A promising technology to increase battery performance and safety are all-solid-state batteries (ASSBs) with a solid Li-ion conducting electrolyte (SE). ASSBs can potentially enable Li-metal anodes, significantly increasing the achievable volumetric and gravimetric energy density. \cite{janek2016solid} However, ASSBs still face several limitations. These include stability issues, high charge transfer resistances at the numerous solid-solid interfaces, insufficient ionic conductivity of the SE, and non-optimized cathode design. \cite{janek2023challenges,ren2023oxide}

The energy and power density of an ASSB are strongly influenced by the cathode, which is typically a composite structure composed of SE and cathode active material (CAM) particles, forming interconnected clusters. \cite{minnmann2022designing} 
For superior performance, both the effective ionic conductivity in the SE phase and the electronic conductivity in the CAM phase must be sufficiently high. Furthermore, achieving a high energy density necessitates thick cathodes and substantial CAM loading. \cite{janek2023challenges}
However, increased CAM fractions lead to increased effective tortuosity\footnote{In this article, we generally refer to the effective tortuosity. For an overview of the various types of tortuosity and their respective definitions, see Ref. \cite{holzer2023tortuosity}.} in the SE phase and low effective ionic conductivity. \cite{minnmann2021editors} This is critical at high cathode thickness and elevated current densities, where ionic limitations can lead to poor CAM utilization. \cite{finsterbusch2018high} So far, enabling both high energy and power density in ASSBs is prevented by either low CAM loading or insufficient effective ionic conductivity of the composite cathode. \cite{nam2018toward,zhang2017interfacial}

An essential requirement for high effective ionic conductivity in energy-dense cathodes is a high bulk conductivity of the SE. \cite{janek2023challenges} Furthermore, the microstructure significantly impacts charge transfer within the composite cathode. Recently, experimental and simulation studies have focused on optimizing composite cathode microstructure for improved cell performance. \cite{clausnitzer2023optimizing,minnmann2021editors,finsterbusch2018high,neumann2021effect,nam2018toward,zhang2017interfacial,strauss2018impact,kaiser2018ion,bielefeld2018microstructural,bielefeld2020modeling,bielefeld2022influence} 

Ionic conductivity in the cathode can be increased by reducing effective tortuosity in the SE phase. An effective way for reducing tortuosity in the SE is to increase SE fractions in the cathode. \cite{minnmann2021editors,kaiser2018ion,clausnitzer2023optimizing} However, this comes at the cost of diminished CAM fractions, resulting in a reduced energy density.
Moreover, cathode void volume must be minimized, as voids lead to limitations of ionic transport, especially at elevated CAM fractions. \cite{clausnitzer2023optimizing,froboese2019effect}
Tortuosity in the SE phase is also influenced by the size of SE and CAM particles. Minimum effective tortuosities are achieved for large CAM and small SE particles. \cite{shi2020high,cronau2022ionic,minnmann2022designing} However, a high ratio between CAM and SE particle size can lead to transport limitations due to a higher number of grain boundaries in the SE and longer diffusion pathways in the CAM. \cite{clausnitzer2023optimizing,yu2017grain,ates2022elucidating,neumann2021effect} 

 These considerations impose inherent trade-offs, limiting the potential of structural optimization of homogeneous composite cathodes for improved cell performance. However, electrode structuring techniques can address some constraints in homogeneous cathodes. Introducing specific inhomogeneities in the microstructure can improve the effective ionic conductivity while maintaining high CAM loading, aiming for both high power and energy density. However, the development of such concepts for ASSBs has been rarely reported in the literature. 

In recent years, perforated electrodes have emerged as a promising strategy to enhance charge transfer in conventional LIBs. \cite{smyrek2015laser,mangang2016influence,kim2018improving,habedank2018increasing,habedank2019enhanced,kraft2020modeling,chen2020efficient,kriegler2021enhanced,de2021beneficial,goel2023optimization}
Perforations in the electrodes, typically induced via laser processing, are infiltrated by the liquid electrolyte (LE), providing direct channels for fast ionic transport through the electrode. Perforated electrodes effectively reduce concentration gradients, improving ionic transport and thus enabling better cell performance at elevated currents. \cite{de2021beneficial} Similarly, multilayer coatings were suggested to enhance ion transport in the porous electrodes, providing higher power density. \cite{gottschalk2023improving,chen2016improvement,wood2021impact,cheng2020combining} Transfer of these approaches to ASSBs might be a viable pathway towards improved cell performance.

Recently, Rosen et al. presented a novel layered cathode design, layering three distinct composite compositions using tape casting. \cite{rosen2022free} By increasing the SE fraction at the separator side of the cathode, the ionic transport in the cathode can be improved. At the same time, an increased CAM fraction at the current collector side ensures high CAM loading and good electronic transport.
The layered cathodes showed better cell performance than homogeneous cathode structures at low experimental current densities.  
Bielefeld et al. used structure-resolved simulations to investigate the potential advantages of a cone-like cathode structure for ASSBs. \cite{bielefeld2022influence} In their simplified model geometries, the SE fraction decreases continuously from separator to current collector, leading to a moderate decrease in overpotential compared to an unstructured electrode.   

While there has been increasing interest in structuring techniques for ASSBs, a comprehensive study is still missing in the literature. This work uses structure-resolved simulations to explore the potential advantages of structured cathode designs for ASSBs. We investigate a perforated and two-layer cathode concept. Our simulation approach focuses on the correlations between cathode microstructure and electrochemical cell performance, enabling us to identify optimal configurations. We simulate LIB and ASSB scenarios to show the material-dependent requirements for an optimum microstructure. With our physics-based simulation approach, we aim to provide guidelines for future developments.

\section{Simulation Methodology}
\subsection{Simulation Workflow}

\begin{figure}[H]
    \centering
    \includegraphics[width=1\textwidth]{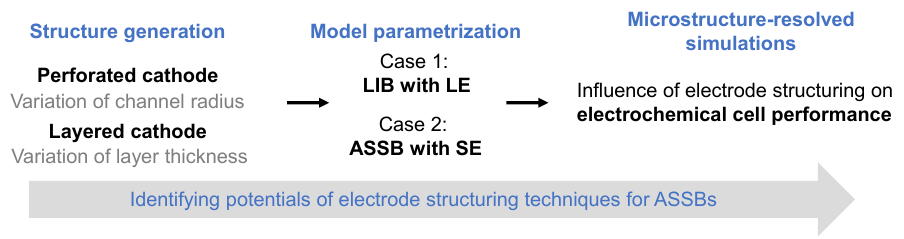}
    \caption{Simulation workflow for investigating the impact of electrode structuring techniques on the electrochemical cell performance for ASSBs.
    }  
    \label{fig:Simulation_workflow}
\end{figure}

We use structure-resolved simulations to explore the influence of electrode structuring techniques on the electrochemical cell performance of ASSBs. Figure \ref{fig:Simulation_workflow} provides an overview of the simulation workflow. Initially, we generate virtual microstructures that serve as the input for our 3D simulations. We focus on two sets of structures: Cells with a perforated and layered cathode design. For the perforated cathodes, we vary the perforation size. For the layered design, we focus on a two-layer concept, adjusting the individual layer thickness. 
This approach allows us to identify optimum configurations for specific operating conditions. Additionally, we simulate the corresponding structuring concepts in LIBs, serving as a reference. By evaluating and comparing our simulation results, we identify limiting processes and show the potential benefits of electrode structuring techniques for ASSBs.

\subsection{Structure generation}
\label{sec:structure_generation}
For this study, we generate virtual microstructures of a perforated and layered cathode design. Details on the stochastic 3D structure generator, calibrated to the microstructure of conventional LIB cathodes, for homogeneous electrodes with different electrode density and particle size are provided in Ref.\cite{knorr2022}. 
To facilitate a direct comparison between LIBs and ASSBs, we maintain a consistent cathode structure in both cases by using the generated structures for LIB and ASSB simulations. However, for the ASSB simulations, the CBD phase in the structures is fully substituted by SE, representing binder-free electrode concepts. \cite{yamamoto2018binder,hippauf2019overcoming}
In all cases, the area-specific theoretical capacity of our reference structure is approximately 7 mAh/cm$^2$, and the structuring concepts reduce the theoretical capacity.
The voxel-based structures are used as input for our electrochemical model, enabling direct correlation between microstructure and electrochemical cell performance.

\paragraph{Perforated cathodes}

Perforated electrodes are commonly manufactured with a symmetric pattern of holes throughout the electrode. In LIBs, the LE wets the channels, facilitating fast ion transport. Due to the symmetric hole pattern, these structures can be efficiently modeled using representative geometries that consider quarter holes and apply isolating boundary conditions. \cite{de2021beneficial,chen2020efficient,habedank2018increasing}. This approach is depicted in Figure S1 (a). 

\begin{figure}[H]
    \centering
    \includegraphics[width=0.8\textwidth]{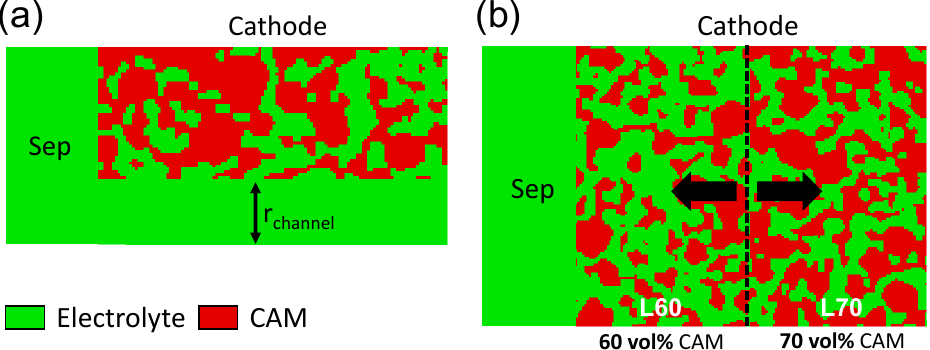}
    \caption{Generated cathode microstructures for the ASSB case. (a) Cross-sectional view of a perforated cathode with a variable channel radius between 0 and 36 $\mu m$. (b) Cross-sectional view of a layered cathode with a first layer containing 60 vol\% CAM at the separator side, and a second layer with 70 vol\% CAM at the current collector side. While the overall thickness of the cathode is held constant, individual layer thickness is varied.}  
    \label{fig:Structure_generation}
\end{figure}

We use a homogeneous cathode microstructure with a CAM fraction of 65 vol\%, an electrolyte fraction of 14 vol\%, and a carbon binder domain (CBD) fraction of 21 vol\%. The generated structure has a size of $ 106 \times 60 \times 60$ $\mu m$ (x/y/z), with the x-axis oriented towards the current collector.
We then introduce electrolyte channels into the structure by substituting CAM and CBD voxels with electrolyte up to a specific radius, simulating the desired perforation pattern. Dimensions in the y and z directions are the distance between hole centers in this symmetric simulation setup.
Figure \ref{fig:Structure_generation} (a) displays the cross-section of the generated cathode structures, characterized by the channel radius $r_\text{channel}$. We vary $r_\text{channel}$ between 0 and 36 $\mu m$, leading to a reduction of up to 28\% of the CAM fraction relative to the original structure. 

\paragraph{Layered cathodes}

We investigate a two-layer concept as a basic representation of a wider spectrum of structuring strategies, ranging from multi-layer designs to gradient configurations. \cite{rosen2022free} These strategies aim to improve cell performance by increasing the CAM fraction across the cathode length from the separator to the current collector. 
In our two-layer configuration, the first layer L$_\text{60}$ at the separator side of the cathode, has a CAM fraction of $\approx$ 60 vol\%. At the current collector side, the second layer L$_\text{70}$ contains an increased CAM fraction of $\approx$ 70 vol\%. The layer with 70 vol\% CAM possesses a high CAM loading while maintaining a percolating network in the electrolyte phase. 
Reducing the CAM fraction to 60 vol\% (L$_\text{60}$) significantly enhances the effective ionic conductivity at still a substantial CAM fraction. \cite{minnmann2021editors,clausnitzer2023optimizing} To assess the potential of the two-layer design in enhancing cell performance, we adjust the thickness of each layer, denoted by $d_\text{L60}$ and $d_\text{L70}$, respectively, while maintaining a constant overall cathode thickness.

For the simulation study in the present paper, we vary the layer thickness fraction $f_\text{L60}$, which represents the ratio of the thickness of layer L$_\text{60}$ to the overall cathode thickness: 

\begin{equation}
    f_\text{L60}=\frac{d_\text{L60}} {d_\text{L60}+d_\text{L70}}
    \label{equ:f_L1}
\end{equation}

We generate the layered structures starting from two homogeneous structures with approximately 60 and 70 vol\% CAM that serve as reference points for our simulations. From these structures, we derive the layers L$_\text{60}$ and L$_\text{70}$, which are subsequently stacked based on the specific configuration.
Figure \ref{fig:Structure_generation} (b) illustrates our generated structures in a cross-section, while Figure S1 (b) shows an exemplary 3D structure. The size of the generated structures is $100 \times 80 \times 80$ $\mu m$. 

\paragraph{Virtual cell assembly}

To generate the input geometry for our simulations, we add a planar anode, separator, and current collectors to the cathode structures. An overview of the simulation geometries is shown in Figure S2.

\subsection{Simulation framework}
\label{sec:simulation_framework}

For this study, we employ the Battery and Electrochemistry Simulation Tool (BEST), a finite volume implementation developed at DLR and Fraunhofer ITWM. \cite{bestwebsite} Within the simulation framework, the charge transport in the battery cell is calculated based on a set of coupled partial differential equations derived from the conservation equations for mass and charge.\cite{latz2011thermodynamic,latz2015multiscale, clausnitzer2023optimizing} Table \ref{tab:Equations_BEST} provides an overview of the governing equations. 

In LEs, Li-ions are transported through migration and diffusion. Migration-based transport is driven by gradients in the electric field. The fraction of the resulting current carried by Li-ions is characterized by the transference number $t_\text{Li}^+$. \cite{newman2021electrochemical} For LEs, $t_\text{Li}^+$ is typically well below 1, suggesting that a substantial part of the current is carried by counter-ions moving opposite the Li-ions. As a result, concentration gradients develop in the electrolyte, leading to Li-ion transport by diffusion. In contrast, most SEs are considered single-ion conductors with $t_\text{Li}^+$=1. In this case, the system of equations in the electrolyte reduces to the Poisson equation with constant concentration within the electrolyte.

\begin{table}
    \centering
    \begin{tabularx}{14cm}{ll}
\toprule
   Equation & Short description\\
   \midrule
   \textbf{Transport in active material}\\
    $\frac{\partial c_\text{AM}}{\partial t}=-\nabla \cdot \left( -D_\text{AM} \nabla c_\text{AM} \right)$ & Mass balance \\
    $0=-\nabla \cdot i_\text{AM}$ & Charge balance \\
    $i_\text{AM}=-\sigma_\text{AM} \nabla \Phi_\text{AM}$ & Electric current \\
\midrule    
   \textbf{Transport in electrolyte}\\
   \textbf{LE ($t_\text{Li}^\text{+}<1$)} \\
    $\frac{\partial c_\text{e}}{\partial t}=-\nabla \cdot \left( -D_\text{e} \nabla c_\text{e}+ \frac{t_\text{Li}^\text{+} \cdot i_\text{e}}{F} \right)$ & Mass balance \\
    $0=-\nabla \cdot i_\text{e}$ & Charge balance \\
    $i_\text{e}=-\kappa \nabla \Phi_\text{e}-\kappa_\text{D} \nabla c_\text{e}$ & Ionic current \\
    $\kappa_\text{D}=\frac{\kappa (t_\text{Li}^\text{+}-1)}{F} \left( \frac{\partial \mu_e}{\partial c_e} \right)$ & \\
    
    \textbf{SE ($t_\text{Li}^\text{+}=1$)}\\
    $0=-\nabla \cdot i_\text{e}$ & Charge balance \\
    $i_\text{e}=-\sigma_\text{Li}^\text{e} \nabla \Phi_\text{e}$ & Ionic current \\
\midrule
   \textbf{Interface between AM and electrolyte}\\
    $i_\text{BV}=i_\text{0} \left[ \exp{\left(\frac{\alpha F}{RT} \eta \right)}-\exp{\left(-\frac{\left( 1-\alpha \right)F}{RT}\eta\right)}\right] $ & Butler-Volmer current  \\
    $i_\text{0}=i_\text{00}^\text{AM} c_\text{e}^\alpha c_\text{AM}^\alpha \left( c_\text{AM}^\text{max}-c_\text{AM}\right)^{1-\alpha}$ & Exchange current density  \\   
\bottomrule
\end{tabularx}
    \caption{Governing equations used in BEST.\cite{latz2011thermodynamic,latz2015multiscale,danner2016thick} Transport equations for the electrolyte phase simplify when assuming a SE with transference number of $t_\text{Li}^\text{+}=1$.}
    \label{tab:Equations_BEST}
\end{table}

\subsection{Material parameters}

We simulate a LIB case with LE and an ASSB case with SE. The respective material parameters are taken from the literature. Table S1 gives an overview of the parameters and corresponding references. Any deviations from these parameters are specified in the relevant section.

\paragraph{NMC811}
In our simulations we consider the cathode active material \ce{LiNi_{0.8}Mn_{0.1}Co_{0.1}O_2} (NMC811). Material-specific parameters such as open circuit voltage, diffusion coefficient, and electric conductivity depend on the lithiation state and are included as functional parameters. \cite{bielefeld2022influence,amin2016characterization,ruess2020influence}

In LIBs, microcracks that develop in NMC811 particles during cycling are invaded by the LE. \cite{ryu2018capacity} This results in shorter diffusion pathways in the CAM and increased active surface area. Consequently, the effective diffusion coefficient and charge transfer kinetics are higher compared to SEs. \cite{ruess2020influence} 

\paragraph{Liquid electrolyte}
In our LIB simulations, we consider a LE with an initial concentration of 1M \ce{LiPF_6} in EC:EMC(3:7). The respective material parameters are taken from the literature.\cite{nyman2008electrochemical,landesfeind2019temperature} Ionic conductivity, diffusion coefficient, transference number, and thermodynamic factor are included as concentration-dependent parameters. In our simulations, we consider both the separator and CBD as homogenized media, characterized by effective transport parameters. \cite{knorr2022} The porous separator with a porosity of 50\% is completely soaked with the LE. The effective conductivity is 50\% of the bulk conductivity. Additionally, our input geometries include a CBD phase with 50\% porosity \cite{vierrath2015morphology,zielke2015three} and an effective ionic conductivity of 12\% of the bulk conductivity.

\paragraph{Solid electrolyte}
For the SE, we consider the argyrodite \ce{Li_6PS_5Cl}. The material parameters are taken from the literature. \cite{ruess2020influence} Contrary to the LIB scenario, the separator is assumed to completely consist of SE. In particular, we consider a binder-free electrode. 

\section{Results and Discussion}

This study uses structure-resolved simulations to determine the potential of a perforated and a two-layer cathode design for improved cell performance. Relevant performance indicators, such as capacity or energy density, are defined in the supporting material. For each design strategy, we provide an overview of the impact of the cathode structuring on theoretical capacity and ionic conductivity. From our simulation results, we identify limiting processes and optimal structures.

\subsection{Perforated cathodes}
\label{sec:perforated_cathodes}
\subsubsection{Overview}

As described in Section \ref{sec:structure_generation}, we generate perforated structures with varying channel radius between 0 and 36 $\mu m$. These perforations are completely filled with electrolyte, providing channels for ionic transport. Increasing the channel radius leads to reduced effective tortuosity in the electrolyte phase and higher effective ionic conductivity. However, the CAM fraction in the cathode is reduced, resulting in a decrease in theoretical capacity.

\begin{figure}[H]
    \centering
    \includegraphics[width=0.8\textwidth]{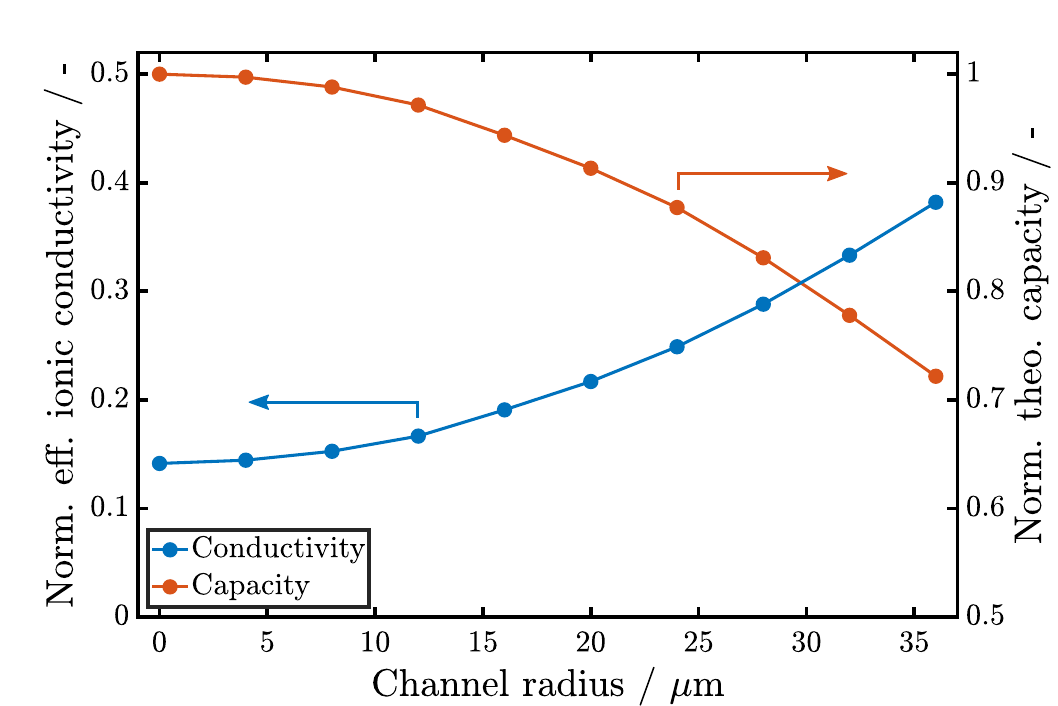}
    \caption{Effect of channel radius on effective ionic conductivity and theoretical capacity. The effective ionic conductivity is normalized with the bulk conductivity of the electrolyte. The theoretical capacity is normalized by the theoretical capacity of the structure without perforation.}  
    \label{fig:Cond_perforation}
\end{figure}

Figure \ref{fig:Cond_perforation} highlights the influence of the channel radius on both normalized effective ionic conductivity and normalized theoretical capacity. As the channel radius increases, the theoretical capacity drops significantly. For the largest channel radius, it reduces to 72\% of the capacity of the unstructured cathode. However, larger channel diameters significantly increase effective ionic conductivity, owing to the reduced effective tortuosity in the electrolyte phase. Therefore, ionic transport limitations are mitigated.

\subsubsection{Electrochemical cell performance}
\paragraph{LIB case}

\begin{figure}[H]
    \centering
    \includegraphics[width=1\textwidth]{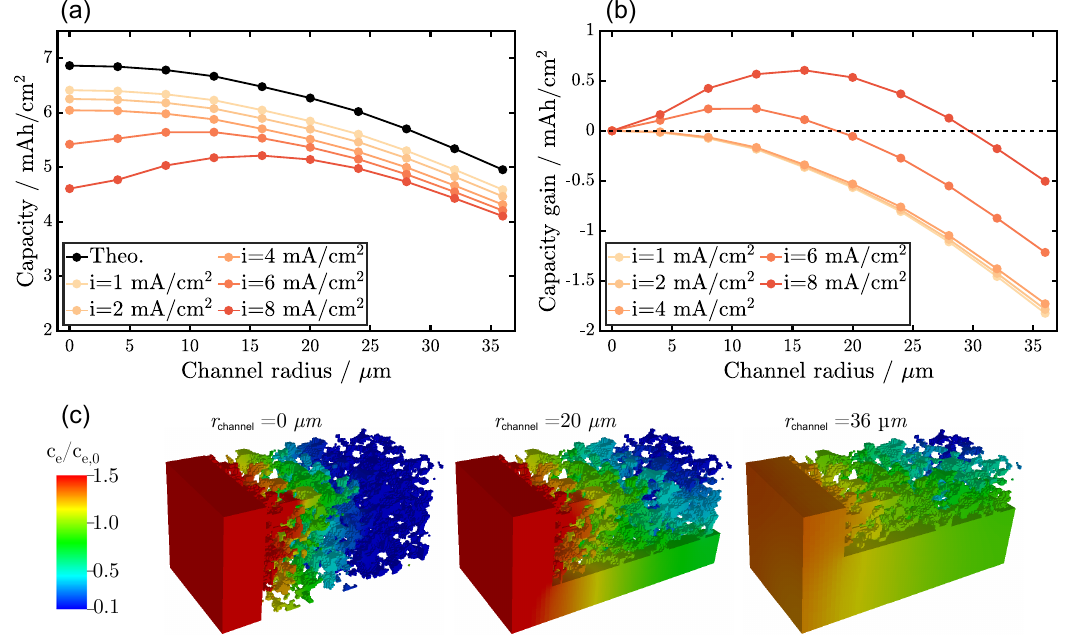}
    \caption{Effect of channel radius on electrochemical cell performance for the LIB case. (a) Practical capacity for various current densities. The black line represents the theoretical capacity of structures with channels. (b) Capacity gain of perforated structures at varying current density compared to the non-perforated structure ($r_\text{channel}=0$ $\mu m$). (c) Normalized Li-ion concentration in the electrolyte phase within the separator and cathode for increasing channel radius at 8 mA/cm$^2$.}
    \label{fig:Perforation_LE}
\end{figure}

We first conduct simulations for a LIB with LE to demonstrate the potential advantages of a perforated cathode design. Figure \ref{fig:Perforation_LE} (a) shows the simulated capacities for varying channel radius at current densities between 1 and 8 mA/cm$^2$. The theoretical capacity, depicted with the black curve, decreases with increasing channel radius due to the lower CAM loading. Due to transport limitations, elevated current densities lead to higher overpotentials and decreasing capacities. However, larger electrolyte channels mitigate these transport limitations. At higher current densities, these channels have an increasingly positive effect on capacity.

Figure \ref{fig:Perforation_LE} (b) shows the capacity gain of the cathodes with channels compared to the homogeneous reference ($r_\text{channel}=0$ $\mu m$). At low current densities, kinetic limitations are minimal, resulting in almost full CAM utilization. Due to the lower theoretical capacities, larger channels lead to decreasing capacities. However, as current density increases, kinetic limitations become more significant. Insufficient ionic transport can cause reduced CAM utilization across the cathode thickness, posing a considerable challenge for high-energy-density cathodes. \cite{singh2015thick,danner2016thick} Significant concentration gradients can develop within the electrolyte, causing concentration overpotentials and charge transport limitations. 
The improved ionic transport due to the channels in the electrodes effectively reduces concentration gradients in the electrolyte, enhancing CAM utilization across the cathode length.\cite{de2021beneficial} At the highest simulated current ($i=8$ mA/cm$^2$), the capacity reaches its maximum value for a channel radius of 16 $\mu m$, with a significant increase of 0.6 mAh/cm$^2$ compared to the homogeneous electrode.

Figure \ref{fig:Perforation_LE} (c) shows the Li-ion concentration in the electrolyte phase for channel radii of 0, 20, and 36 $\mu m$. As channel radius increases, concentration gradients decrease significantly across the cathode length, leading to more efficient charge transport in the electrolyte and reduced concentration overpotential.

\begin{figure}[H]
    \centering
    \includegraphics[width=1\textwidth]{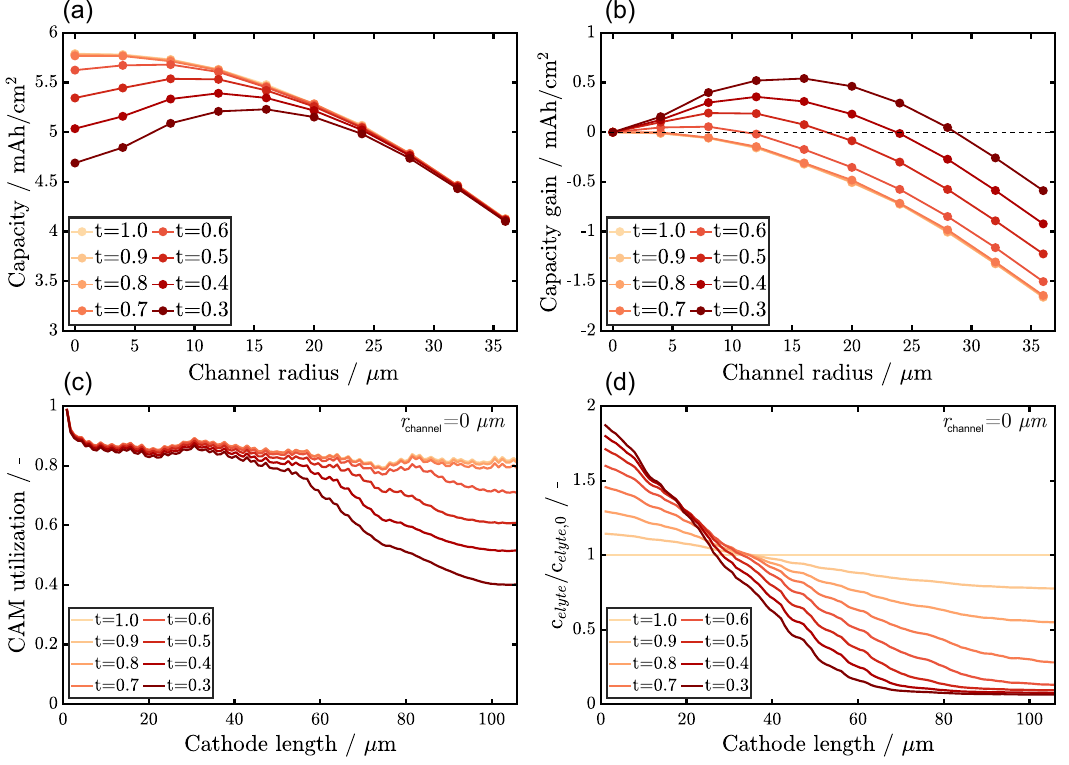}
    \caption{Effect of increasing transference number on electrochemical cell performance for perforated cathodes assuming parameters of the LIB case. (a) Impact of transference number on practical capacity at 8 mA/cm$^2$. (b) Impact of transference number on capacity gain compared to the non-perforated structure ($r_\text{channel}=0$ $\mu m$) at 8 mA/cm$^2$. (c) Impact of transference number on the mean CAM utilization across the cathode thickness from separator to current collector for $r_\text{channel}=0$ $\mu m$ at 8 mA/cm$^2$. (d) Impact of transference number on the mean Li-ion concentration in the electrolyte across the cathode thickness from separator to current collector for $r_\text{channel}=0$ $\mu m$ at 8 mA/cm$^2$. The Li-ion concentration in the electrolyte is normalized by the initial Li-ion concentration. }  
    \label{fig:Perforation_t}
\end{figure}

\paragraph{Effect of transference number}
The transference number of the electrolyte provides insight into the relative contributions of diffusion and migration to the overall ionic transport. A low transference number indicates diffusion-dominated Li-ion transport, while a high transference number indicates migration-dominated transport. Addressing the need for efficient charge transport in the electrolyte and minimizing concentration overpotentials has led to intensive efforts in finding electrolytes with both high conductivity and transference number. \cite{diederichsen2017promising,logan2020electrolyte,zhou2023strategies} Inorganic SEs with high lithium conductivity and transference number are represented by transference numbers close to 1.
Figure \ref{fig:Perforation_t} illustrates the impact of increasing transference number on cell performance for electrodes with electrolyte channels at 8 mA/cm$^2$.

Figure \ref{fig:Perforation_t} (a) shows the simulated capacities for transference numbers between 0.3 and 1 at 8 mA/cm$^2$. As discussed in the previous section, structures with smaller channel radii exhibit high concentration gradients in the electrolyte, leading to significant concentration overpotentials. For these configurations, an increase in the transference number results in a significant capacity gain. However, ionic transport limitations are less pronounced for cathodes with larger channels, reducing the influence of transference number on cell performance.

As $t_\text{Li}^\text{+}$ increases, capacities rise due to improved ionic transport. Concentration gradients across the cathode length diminish with increasing transference number (Figure \ref{fig:Perforation_t} (d)). The more efficient ionic transport has a direct impact on CAM utilization throughout the cathode (Figure \ref{fig:Perforation_t} (c)). While at $t_\text{Li}^\text{+}=0.3$, the CAM near the current collector is less utilized, at $t_\text{Li}^\text{+}=1$, utilization across the cathode length is almost constant.

Figure \ref{fig:Perforation_t} (b) shows the capacity gain due to electrolyte channels compared to the homogeneous electrode for varying transference numbers. As the transference number increases, concentration gradients in the electrolyte are reduced, diminishing capacity gains due to electrolyte channels.

 At high transference numbers ($t_\text{Li}^\text{+}>0.7$), the perforated structures consistently show lower capacities than the non-perforated electrode at 8 mA/cm$^2$. On the one hand, this shows that the predominant benefit of electrolyte channels is enhanced diffusive transport, while the impact on migration is relatively minor. On the other hand, the results demonstrate that SEs with comparable ionic conductivity to LEs could pave the way toward efficient, high-performance cells with high power and energy density.

\paragraph{ASSB case}

In the next step, we specifically focus on the influence of perforated cathode structures on ASSB performance with state-of-the-art SEs. We employ the SE parameters outlined in Table S1.

\begin{figure}[H]
    \centering
    \includegraphics[width=1\textwidth]{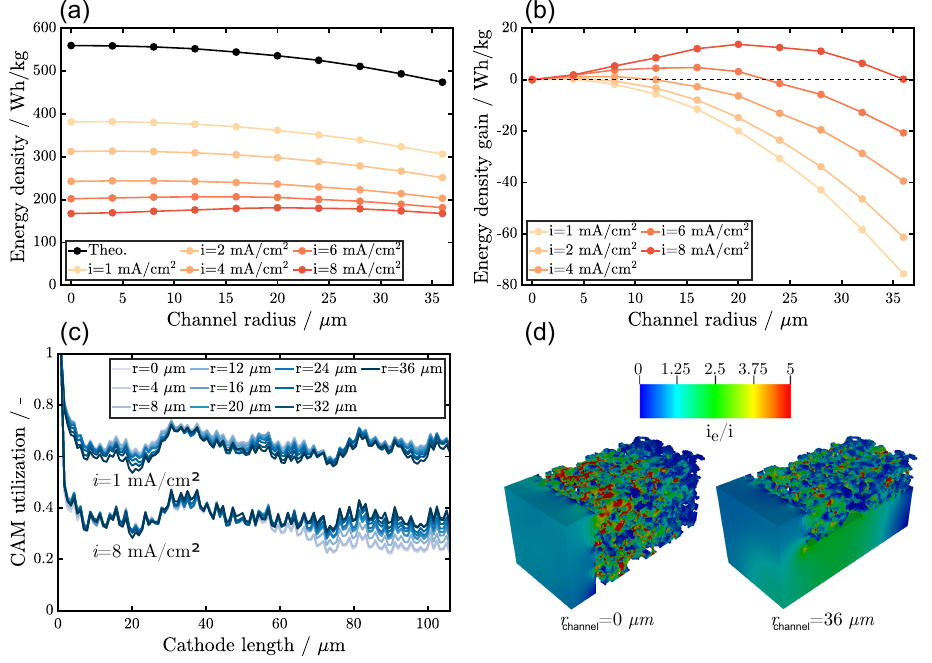}
    \caption{Effect of channel radius on electrochemical cell performance for the ASSB case. (a) Energy density for various current densities. The black line represents the theoretical energy density of the perforated structures. (b) Energy density gain of structures with channels at different current densities compared to the non-perforated structure ($r_\text{channel}=0$ $\mu m$). (c) Impact of channel radius on the mean CAM utilization across the cathode thickness from separator to current collector at 1 mA/cm$^2$ and 8 mA/cm$^2$. (d) Current distribution in the electrolyte phase at the end of discharge for $r_\text{channel}=0$ $\mu m$ and $r_\text{channel}=36$ $\mu m$ and 8 mA/cm$^2$. Red regions indicate hot spots with high current density.}  
    \label{fig:Perforation_SE}
\end{figure}

The primary goal of employing electrode structuring techniques is maximizing energy density at elevated current densities. Figure \ref{fig:Perforation_SE} (a) demonstrates the impact of channel size on the energy density for current densities from 1 to 8 mA/cm$^2$. 
With increasing current density, the energy density decreases due to kinetic limitations, resulting in lower utilization of the CAM. Figure S3 shows the simulation results in terms of capacity. In contrast to the LIB case (cf. Figure \ref{fig:Perforation_LE}), the cell performance is notably below the theoretical values even at low current densities. This is caused by the specific material parameters for the ASSB case. The SE has lower ionic conductivity compared to the LE. Additionally, we consider a reduced effective CAM diffusivity and exchange current density at the SE/CAM interface. Thus, despite the high transference number of the SE, CAM utilization is significantly lower than in the LIB case. 

Figure \ref{fig:Perforation_SE} (b) displays the energy density gain of cathodes with electrolyte channels relative to the homogeneous electrode. It is important to note that for the material system \ce{Li_6PS_5Cl}/NMC811, the SE has a significantly lower density than the CAM. The reduced mass of structures containing more SE positively impacts energy density. Therefore, the calculated energy densities for cathodes with larger channels show a more favorable trend compared to the capacities depicted in Figure S3.

As discussed in the previous section, channels do not significantly enhance migration-dominated Li-ion transport. Still, they provide shorter conduction pathways and reduced effective tortuosity in the SE phase. Therefore, structures with electrolyte channels show higher energy densities at higher current densities than the homogeneous structure. However, the maximum energy density gain at 8 mA/cm$^2$ is modest (14 Wh/kg at $r_\text{channel}=20$ $\mu m$).

Figure \ref{fig:Perforation_SE} (c) displays the CAM utilization across the cathode length depending on channel size. Interestingly, at 1 mA/cm$^2$, CAM utilization decreases for larger channel sizes despite shorter conduction pathways. This can be explained by the reduced active areas in the structures with channels, resulting in larger interfacial currents and higher overpotentials. 
At 8 mA/cm$^2$, the shorter ionic conduction pathways in structures with channels lead to better CAM utilization at the current-collector side of the cathode.
Figure \ref{fig:Perforation_SE} (d) shows the current distribution in the electrolyte phase during discharge for $r_\text{channel}=0$ $\mu m$ and $r_\text{channel}=36$ $\mu m$. Larger channels provide shorter ionic conduction pathways, resulting in lower peak currents and fewer hot spots. 

Our simulations indicate that the potential of using cathodes with channels to improve ASSB performance is quite limited. Given the technical and economic challenges associated with manufacturing such electrodes for ASSBs, other concepts might be more promising. 

\subsection{Layered cathodes}
\subsubsection{Overview}
As a second concept, we investigate a two-layer cathode design. The layered structures have a reduced CAM fraction (60 vol\%) near the separator, enhancing ionic transport. Simultaneously, the layer close to the current collector has an increased CAM fraction (70 vol\%), ensuring high CAM loading and enhanced effective electronic conductivity.

In our geometries, we vary the relative thickness of the separator layer $f_\text{L60}$, maintaining a consistent overall cathode thickness (cf. Section \ref{sec:structure_generation}). For reference, $f_\text{L60}=0$ corresponds to a homogeneous cathode structure with 70 vol\% CAM. In contrast, $f_\text{L60}=1$ characterizes a homogeneous cathode with 60 vol\% CAM.

\begin{figure}[H]
    \centering
    \includegraphics[width=0.8\textwidth]{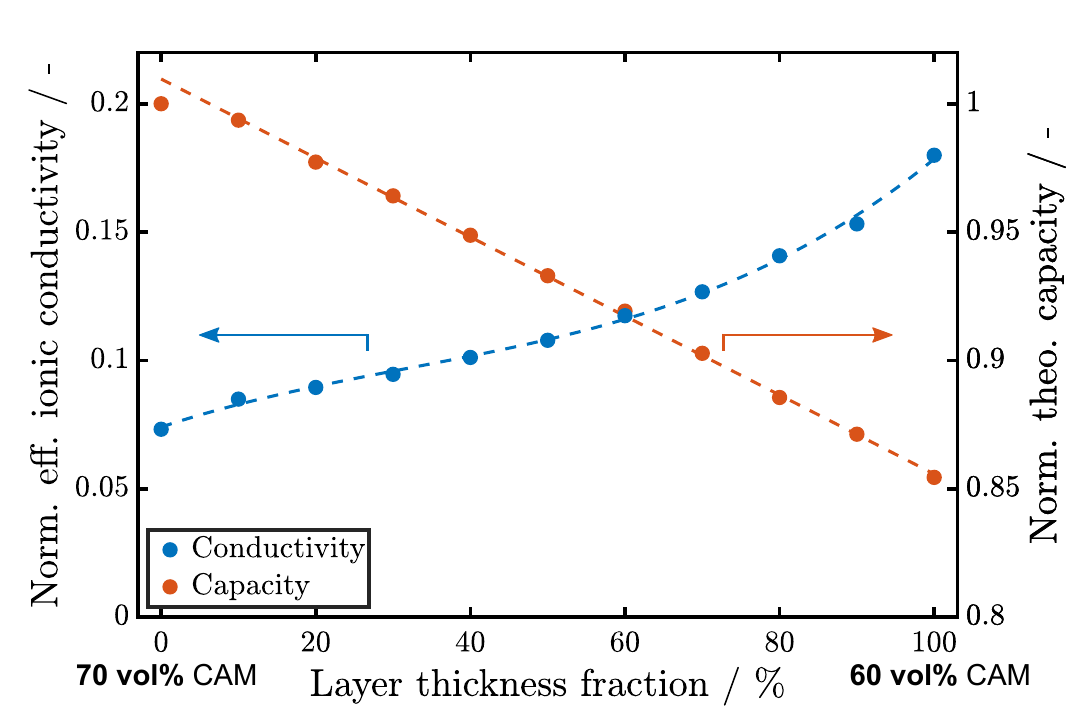}
    \caption{Impact of layer thickness fraction $f_\text{L60}$ on effective ionic conductivity and theoretical capacity for a two-layer concept. The first layer adjacent to the separator has a CAM volume fraction of 60 vol\%, while the second layer near the current collector has a CAM volume fraction of 70 vol\%. The effective ionic conductivity is normalized with the bulk conductivity of the electrolyte. The theoretical capacity is normalized with the capacity of the homogeneous structure with 70 vol\% CAM ($f_\text{L60}=0$).}  
    \label{fig:Cond_layered}
\end{figure}

Figure \ref{fig:Cond_layered} shows the impact of increasing $f_\text{L60}$ on effective ionic conductivity and theoretical capacity. As we extend layer L60, the effective ionic conductivity increases. However, the lower CAM fraction in the structures results in a decrease in theoretical capacity. 

Unlike the capacity, the effective ionic conductivity does not follow a linear trend. For larger $f_\text{L60}$, the increase in effective ionic conductivity becomes more pronounced.

\subsubsection{Electrochemical cell performance}

\begin{figure}[H]
    \centering
    \includegraphics[width=1\textwidth]{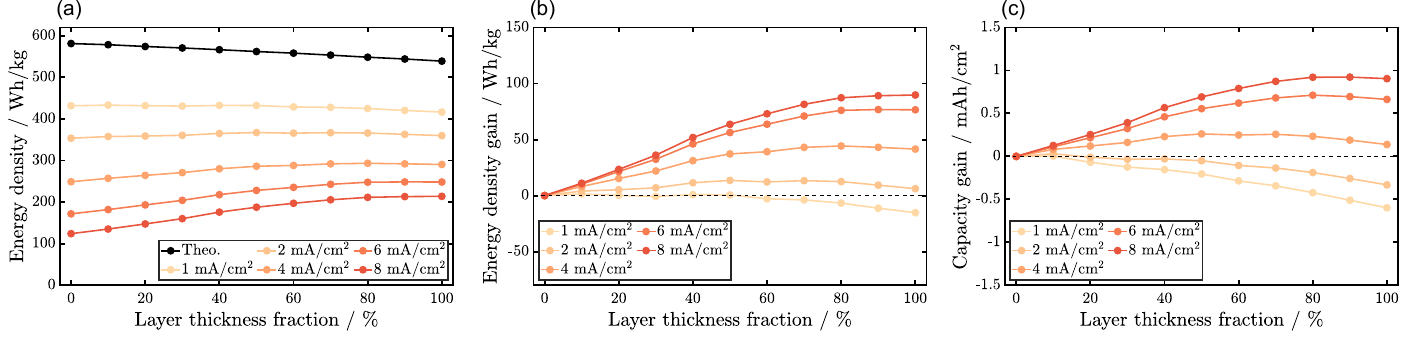}
    \caption{Influence of layer thickness fraction of the two-layered cathodes on electrochemical ASSB performance. (a) Energy density for current densities ranging from 1 to 8 mA/cm$^2$. The black line represents the theoretical energy density of the generated structures. (b) Energy density gain realized for the layered structures compared to a homogeneous cathode structure with 70\% CAM ($f_\text{L60}=0\%$). (c) Corresponding capacity gain.}  
    \label{fig:Layered_SE}
\end{figure}

Figure \ref{fig:Layered_SE} (a) illustrates the influence of the two-layer design on ASSB cell performance. Due to kinetic constraints, the simulated energy densities are significantly lower than the theoretical values, even at the smallest current density (1 mA/cm$^2$). Losses become more pronounced as the current density increases. The corresponding capacities are presented in Figure S4.

Figure \ref{fig:Layered_SE} shows the gain in energy density (b)  and capacity (c) resulting from a layer with higher SE content close to the current collector. At low current density (1 mA/cm$^2$), the capacity decreases with increasing $f_\text{L60}$ owing to reduced theoretical capacity. This trend is not as pronounced in the energy density due to the lower specific mass of the SE.

At high current densities, the improved ionic transport in layered structures is more prominent, resulting in significantly improved cell performance compared to the non-layered structure with 70 vol\% CAM. As the current density increases, the optimum configuration, balancing ionic transport and CAM loading, is shifted towards a higher thickness of the layer adjacent to the separator. At 4 mA/cm$^2$, the highest capacity is reached at $f_\text{L60}=0.5$. At 6 and 8 mA/cm$^2$, maximum capacities are achieved at  $f_\text{L60}=0.8$ and $f_\text{L60}=0.9$, respectively. 

Given the lower density of \ce{Li_6PS_5Cl} compared to NMC811, maximum energy density is reached at higher $f_\text{L60}$. At elevated current density (6 and 8 mA/cm$^2$), layered cathodes with $f_\text{L60}>0.5$ can attain a gain in energy density exceeding 50 Wh/kg. 

It is important to highlight that the orientation of the layers is critical for optimizing cell performance. Although a reverse structure, with a high CAM fraction near the separator and a low CAM fraction towards the current collector, has the same average properties, cell performance will be worse due to the high effective tortuosity near the separator. Figure S5 illustrates the simulated capacities for the reverse two-layer structures. The simulated capacities are significantly lower, especially at low and moderate $f_\text{L60}$.  

\begin{figure}[H]
    \centering
    \includegraphics[width=1\textwidth]{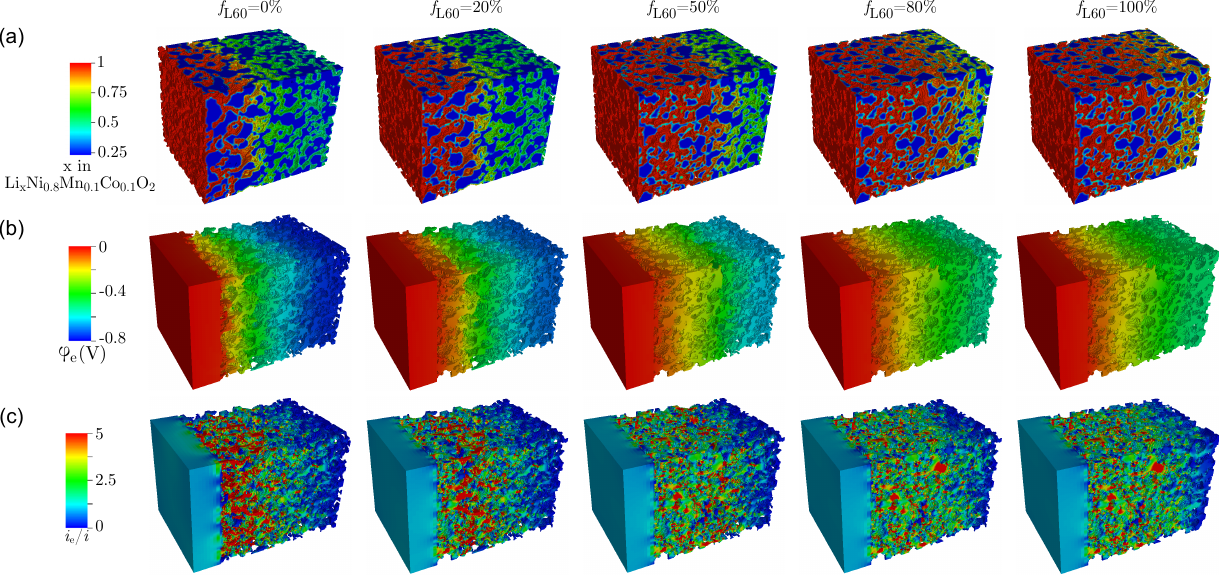}
    \caption{3D simulation results for the two-layered structures with varying $f_\text{L60}$ at the the lower cut-off voltage at $i=8$ mA/cm$^2$. (a) Influence of layer thickness fraction on CAM lithiation. (b) Influence of layer thickness fraction on electrochemical potential of Li-ions in the SE. (c) Influence of layer thickness fraction on ionic current in SE.}  
    \label{fig:Layered_SE_3D}
\end{figure}

Figure \ref{fig:Layered_SE_3D} provides insight into the effect of electrode layers on the charge transfer at $i=8$ mA/cm$^2$.

Figure \ref{fig:Layered_SE_3D} (a) shows the CAM lithiation at the lower cut-off voltage for structures with $f_\text{L60}=0$, 0.2, 0.5, 0.8, and 1. At $f_\text{L60}=0$, high CAM lithiation is primarily observed near the separator. Due to insufficient ionic transport in the SE phase, CAM utilization decreases significantly across the cathode length. 
As $f_\text{L60}$ increases, the charge transport close to the separator significantly improves, resulting in fewer lithiation gradients. In the homogeneous structure with $f_\text{L60}$=1, the enhanced ionic transport enables almost constant lithiation across the cathode length. The generally low CAM utilization at 8 mA/cm$^2$ can be primarily attributed to the low Li mobility in NMC811, resulting in very low lithiation states in the particle centers.

Figure \ref{fig:Layered_SE_3D} (b) demonstrates the effect of increasing $f_\text{L60}$ on ionic transport represented by the electrochemical potential of Li-ions in the electrolyte. Due to lower ionic resistance, structures with high $f_\text{L60}$ show reduced potential gradients in the SE phase across the cathode length, improving cell performance. 

Enhanced ionic transport also reduces maximum local currents, as depicted in Figure \ref{fig:Layered_SE_3D} (c). Practically, minimizing local currents and overpotentials are reported to mitigate local degradation phenomena.\cite{jung2014understanding} For the layered configurations, elevated currents are more prominent at the interface between the two layers.

Interestingly, the layered structure with $f_\text{L60}=0.8$ and the homogeneous structure with 60 vol\% CAM ($f_\text{L60}=1$) exhibit similar current and potential distributions. The two-layer structure shows only slightly lower lithiation of the CAM near the current collector while offering higher theoretical capacity. 

\begin{figure}[H]
    \centering
    \includegraphics[width=1\textwidth]{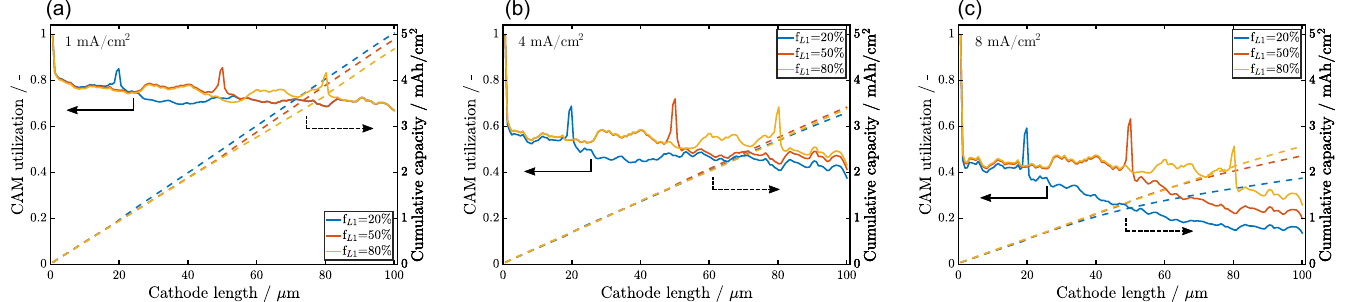}
    \caption{Influence of layer thickness fraction on mean CAM utilization and cumulative capacity across the cathode length. The separator is at 0 $\mu$m and the current collector at 100 $\mu$m. (a) $i=1$ mA/cm$^2$. (b) $i=4$ mA/cm$^2$. (c) $i=8$ mA/cm$^2$.}  
    \label{fig:Layered_SE_vtk}
\end{figure}

Figure \ref{fig:Layered_SE_vtk} shows the CAM utilization and cumulative capacity over the cathode length for layered structures with $f_\text{L60}=0.2$, 0.5, and 0.8.

Across all current densities, a consistent trend in CAM utilization is observed. Generally, layer L60 has better utilization than layer L70 due to its lower effective tortuosity in the SE phase. The charge transfer kinetics govern the average utilization within each layer. CAM utilization drops more noticeably across the cathode length at high current densities due to ionic transport limitations. This decline is more pronounced in layer L70 compared to layer L60.

At the interface between both layers, a distinct peak in CAM utilization is evident. Over a length of approximately 2 $\mu$m, CAM utilization increases noticeably before decreasing to the levels characteristic of layer L70. 
Stacking during structure generation prevents a smooth transition between both layers and yields small CAM particles at the interface between the two layers.
The higher CAM utilization is likely due to the decreased particle size compared to the rest of the structure, providing shorter diffusion pathways. It is worth noting that the peaks in CAM utilization between both layers become more prominent at higher current densities due to enhanced kinetics.
In subsequent studies, applying advanced structure generators might help to ensure a smoother transition between layers L60 and L70. 

At 1 mA/cm$^2$ (Figure \ref{fig:Layered_SE_vtk} (a)), all configurations exhibit consistently high CAM utilization. Given the relatively low current density, the overall capacity depends mainly on the CAM fraction in the structures. Thus, the increase in cumulative capacity is more significant in layer L70 than in layer L60. As a result, the cathode with $f_\text{L60}=0.2$ achieves a greater overall capacity compared to the structures with $f_\text{L60}=0.5$ and $f_\text{L60}=0.8$.

At 4 mA/cm$^2$ (Figure \ref{fig:Layered_SE_vtk} (b)), ionic transport limitations become more pronounced, causing a noticeable decrease in CAM utilization from the separator to the current collector. Additionally, the average CAM utilization in layer L70 is significantly lower than in layer L60. Among the depicted configurations, the structure with $f_\text{L60}=0.8$ has the highest CAM utilization across the entire cathode length. However, despite lower CAM utilization, the cathode with $f_\text{L60}=0.5$ shows slightly higher overall capacity due to the higher CAM loading.

At 8 mA/cm$^2$ (Figure \ref{fig:Layered_SE_vtk} (c)), CAM utilization decreases significantly across the cathode length. This reduction is more significant in layer L70 in comparison to layer L60. Consequently, the cathode with $f_\text{L60}=0.8$ shows significantly higher utilization, which despite its low CAM loading, results in higher capacities compared to the structures with $f_\text{L60}=0.2$ and $f_\text{L60}=0.5$.

Our simulation results show that the optimum configuration for the two-layer cathodes strongly depends on the operating conditions. 
In general, structures are favorable that show only minimal energy density loss at low current density compared to a homogeneous cathode with a high CAM fraction ($f_\text{L60}=0$). Simultaneously, at high current densities, these structures should approach the maximum energy density given by a homogeneous cathode with a low CAM fraction ($f_\text{L60}=1$). Structures with  $f_\text{L60}$ between 0.6 and 0.8 meet these criteria for the material system studied. However, note that the CAM fractions in the two layers are not optimized. Further refinement could enable higher energy densities, possibly favoring different configurations.

\section{Conclusions and Outlook}
The cell performance of ASSBs is limited by the low effective ionic conductivity observed in high-energy-density composite cathodes. Electrode structuring techniques offer the potential to enhance ionic charge transfer in the cathode while still allowing high CAM loading. In this work, we employ structure-resolved simulations to explore the impact of cathode structuring on cell performance. We determine the key factors governing cell performance by analyzing our simulation results for electrode designs with electrolyte channels and layers with varying electrolyte content. Moreover, we are able to identify optimal structures depending on operation conditions.

A promising concept primarily studied for conventional LIBs with liquid electrolytes is the use of electrolyte channels, providing pathways for fast ionic transport. Our simulation results show that channels wetted by a liquid electrolyte can provide significantly increased capacities at elevated current densities. As channel size increases, concentration gradients in the electrolyte decrease, promoting more efficient ionic charge transport. At high current densities, improved transport outweighs the loss of CAM compared to unstructured electrodes. 

However, most inorganic solid electrolytes are single-ion conductors with a transference number close to one. Therefore, concentration gradients in the SE are negligible except for space charge layers close to the CAM interface. As the transference number increases, the primary advantage of electrolyte channels -the reduction in concentration gradients- becomes less significant. Our simulation results show that, despite improved ionic transport through shorter transport pathways, electrolyte channels cannot significantly increase in cathode performance for ASSBs. This emphasizes that electrode designs which are effective for conventional LIBs must be rethought for ASSBs. 

A promising strategy for ASSB development is the use of layered cathodes. Increasing the SE fraction at the separator side and CAM fraction at the current collector side makes it possible to achieve a cathode with high effective ionic and electronic conductivity while maintaining a high CAM loading. 
Our simulations investigate a two-layer concept with 60 vol\% CAM at the separator and 70 vol\% CAM at the current collector. We vary the thickness of both layers, keeping the overall cathode thickness constant. Our simulations indicate a significant potential for improved cell performance using layered cathodes.
Increasing the thickness of the layer with lower CAM content at the separator reduces potential gradients in the electrolyte, resulting in decreased local current densities and overpotentials. Ideally, the capacity loss at low current densities due to reduced CAM fraction is minimal, while the capacity gain at high current densities from enhanced ionic transport is maximized. Depending on the materials and operating conditions, optimal performance was found when the layer at the separator starts to exceed 50\% of the overall thickness. At the highest simulated current density of 8 mA/cm$^2$, the possible energy density achieved by the two-layer concept is approximately 90 Wh/kg. 

Our simulation results encourage experimental studies focusing on layered cathodes for ASSBs. Increasing the number of layers to achieve a smoother gradient in CAM fraction across the cathode length will likely enhance cell performance further. Scalable manufacturing methods like tape-casting can produce multi-layer cathodes, making them particularly interesting for future cell designs. \cite{rosen2022free} Advanced deposition techniques such as powder aerosol deposition (PAD) \cite{nazarenus2021powder} may pave the way for full gradient cathodes.
 
Particle size optimization offers another way to enhance charge transfer within the composite cathode. Similar to the approach explored in this study, the CAM particle size can be varied in the different layers. \cite{usiskin2020guidelines}.
 Further simulation studies might explore a combined approach with gradients in both cathode composition and particle size using novel CAMs specifically designed for application in future ASSBs.

\section*{Supporting Information}

Additional references cited within the Supporting Information. \cite{xu2014lithium,materialsproject}
 
\begin{acknowledgement}

The authors acknowledge support with computational resources provided by the state of Baden-Wurttemberg through bwHPC and the German Research Foundation (DFG) through grant no INST 40/575-1 FUGG (JUSTUS 2 cluster). We gratefully acknowledge the financial support by the German Federal Ministry of Research and Education (BMBF) within the scope of the projects FestBatt2 (FKZ: 03XP0435A) and HiStructures (FKZ: 03XP0243D), as well as from the German Federal Ministry for Economic Affairs and Climate Action (BMWK) in the frame of the project structur.e (FKZ: 03ETE018D). The work of MN was funded by DFG under Project ID 390874152 (POLiS Cluster of Excellence, EXC 2154). The present paper contributes to the research performed at CELEST (Center for Electrochemical Energy Storage Ulm-Karlsruhe).

\end{acknowledgement}

\bibliography{literature.bib}

\providecommand{\latin}[1]{#1}
\makeatletter
\providecommand{\doi}
  {\begingroup\let\do\@makeother\dospecials
  \catcode`\{=1 \catcode`\}=2 \doi@aux}
\providecommand{\doi@aux}[1]{\endgroup\texttt{#1}}
\makeatother
\providecommand*\mcitethebibliography{\thebibliography}
\csname @ifundefined\endcsname{endmcitethebibliography}  {\let\endmcitethebibliography\endthebibliography}{}
\begin{mcitethebibliography}{70}
\providecommand*\natexlab[1]{#1}
\providecommand*\mciteSetBstSublistMode[1]{}
\providecommand*\mciteSetBstMaxWidthForm[2]{}
\providecommand*\mciteBstWouldAddEndPuncttrue
  {\def\EndOfBibitem{\unskip.}}
\providecommand*\mciteBstWouldAddEndPunctfalse
  {\let\EndOfBibitem\relax}
\providecommand*\mciteSetBstMidEndSepPunct[3]{}
\providecommand*\mciteSetBstSublistLabelBeginEnd[3]{}
\providecommand*\EndOfBibitem{}
\mciteSetBstSublistMode{f}
\mciteSetBstMaxWidthForm{subitem}{(\alph{mcitesubitemcount})}
\mciteSetBstSublistLabelBeginEnd
  {\mcitemaxwidthsubitemform\space}
  {\relax}
  {\relax}

\bibitem[Li \latin{et~al.}(2019)Li, Khajepour, and Song]{li2019comprehensive}
Li,~Z.; Khajepour,~A.; Song,~J. A comprehensive review of the key technologies for pure electric vehicles. \emph{Energy} \textbf{2019}, \emph{182}, 824--839\relax
\mciteBstWouldAddEndPuncttrue
\mciteSetBstMidEndSepPunct{\mcitedefaultmidpunct}
{\mcitedefaultendpunct}{\mcitedefaultseppunct}\relax
\EndOfBibitem
\bibitem[Li \latin{et~al.}(2014)Li, Gao, Li, and Yuan]{li2014life}
Li,~B.; Gao,~X.; Li,~J.; Yuan,~C. Life cycle environmental impact of high-capacity lithium ion battery with silicon nanowires anode for electric vehicles. \emph{Environmental Science \& Technology} \textbf{2014}, \emph{48}, 3047--3055\relax
\mciteBstWouldAddEndPuncttrue
\mciteSetBstMidEndSepPunct{\mcitedefaultmidpunct}
{\mcitedefaultendpunct}{\mcitedefaultseppunct}\relax
\EndOfBibitem
\bibitem[Ellingsen \latin{et~al.}(2016)Ellingsen, Singh, and Str{\o}mman]{ellingsen2016size}
Ellingsen,~L. A.-W.; Singh,~B.; Str{\o}mman,~A.~H. The size and range effect: lifecycle greenhouse gas emissions of electric vehicles. \emph{Environmental Research Letters} \textbf{2016}, \emph{11}, 054010\relax
\mciteBstWouldAddEndPuncttrue
\mciteSetBstMidEndSepPunct{\mcitedefaultmidpunct}
{\mcitedefaultendpunct}{\mcitedefaultseppunct}\relax
\EndOfBibitem
\bibitem[Diouf and Pode(2015)Diouf, and Pode]{diouf2015potential}
Diouf,~B.; Pode,~R. Potential of lithium-ion batteries in renewable energy. \emph{Renewable Energy} \textbf{2015}, \emph{76}, 375--380\relax
\mciteBstWouldAddEndPuncttrue
\mciteSetBstMidEndSepPunct{\mcitedefaultmidpunct}
{\mcitedefaultendpunct}{\mcitedefaultseppunct}\relax
\EndOfBibitem
\bibitem[Tu \latin{et~al.}(2019)Tu, Feng, Srdic, and Lukic]{tu2019extreme}
Tu,~H.; Feng,~H.; Srdic,~S.; Lukic,~S. Extreme fast charging of electric vehicles: A technology overview. \emph{IEEE Transactions on Transportation Electrification} \textbf{2019}, \emph{5}, 861--878\relax
\mciteBstWouldAddEndPuncttrue
\mciteSetBstMidEndSepPunct{\mcitedefaultmidpunct}
{\mcitedefaultendpunct}{\mcitedefaultseppunct}\relax
\EndOfBibitem
\bibitem[Li \latin{et~al.}(2020)Li, Feng, Luo, and Chen]{li2020fast}
Li,~M.; Feng,~M.; Luo,~D.; Chen,~Z. Fast charging Li-ion batteries for a new era of electric vehicles. \emph{Cell Reports Physical Science} \textbf{2020}, \emph{1}, 100212\relax
\mciteBstWouldAddEndPuncttrue
\mciteSetBstMidEndSepPunct{\mcitedefaultmidpunct}
{\mcitedefaultendpunct}{\mcitedefaultseppunct}\relax
\EndOfBibitem
\bibitem[Cano \latin{et~al.}(2018)Cano, Banham, Ye, Hintennach, Lu, Fowler, and Chen]{cano2018batteries}
Cano,~Z.~P.; Banham,~D.; Ye,~S.; Hintennach,~A.; Lu,~J.; Fowler,~M.; Chen,~Z. Batteries and fuel cells for emerging electric vehicle markets. \emph{Nature Energy} \textbf{2018}, \emph{3}, 279--289\relax
\mciteBstWouldAddEndPuncttrue
\mciteSetBstMidEndSepPunct{\mcitedefaultmidpunct}
{\mcitedefaultendpunct}{\mcitedefaultseppunct}\relax
\EndOfBibitem
\bibitem[Placke \latin{et~al.}(2017)Placke, Kloepsch, D{\"u}hnen, and Winter]{placke2017lithium}
Placke,~T.; Kloepsch,~R.; D{\"u}hnen,~S.; Winter,~M. Lithium ion, lithium metal, and alternative rechargeable battery technologies: the odyssey for high energy density. \emph{Journal of Solid State Electrochemistry} \textbf{2017}, \emph{21}, 1939--1964\relax
\mciteBstWouldAddEndPuncttrue
\mciteSetBstMidEndSepPunct{\mcitedefaultmidpunct}
{\mcitedefaultendpunct}{\mcitedefaultseppunct}\relax
\EndOfBibitem
\bibitem[Weiss \latin{et~al.}(2021)Weiss, Ruess, Kasnatscheew, Levartovsky, Levy, Minnmann, Stolz, Waldmann, Wohlfahrt-Mehrens, Aurbach, \latin{et~al.} others]{weiss2021fast}
Weiss,~M.; Ruess,~R.; Kasnatscheew,~J.; Levartovsky,~Y.; Levy,~N.~R.; Minnmann,~P.; Stolz,~L.; Waldmann,~T.; Wohlfahrt-Mehrens,~M.; Aurbach,~D.; others Fast charging of lithium-ion batteries: a review of materials aspects. \emph{Advanced Energy Materials} \textbf{2021}, \emph{11}, 2101126\relax
\mciteBstWouldAddEndPuncttrue
\mciteSetBstMidEndSepPunct{\mcitedefaultmidpunct}
{\mcitedefaultendpunct}{\mcitedefaultseppunct}\relax
\EndOfBibitem
\bibitem[Janek and Zeier(2016)Janek, and Zeier]{janek2016solid}
Janek,~J.; Zeier,~W.~G. A solid future for battery development. \emph{Nature Energy} \textbf{2016}, \emph{1}, 1--4\relax
\mciteBstWouldAddEndPuncttrue
\mciteSetBstMidEndSepPunct{\mcitedefaultmidpunct}
{\mcitedefaultendpunct}{\mcitedefaultseppunct}\relax
\EndOfBibitem
\bibitem[Janek and Zeier(2023)Janek, and Zeier]{janek2023challenges}
Janek,~J.; Zeier,~W.~G. Challenges in speeding up solid-state battery development. \emph{Nature Energy} \textbf{2023}, 1--11\relax
\mciteBstWouldAddEndPuncttrue
\mciteSetBstMidEndSepPunct{\mcitedefaultmidpunct}
{\mcitedefaultendpunct}{\mcitedefaultseppunct}\relax
\EndOfBibitem
\bibitem[Ren \latin{et~al.}(2023)Ren, Danner, Moy, Finsterbusch, Hamann, Dippell, Fuchs, M{\"u}ller, Hoft, Weber, \latin{et~al.} others]{ren2023oxide}
Ren,~Y.; Danner,~T.; Moy,~A.; Finsterbusch,~M.; Hamann,~T.; Dippell,~J.; Fuchs,~T.; M{\"u}ller,~M.; Hoft,~R.; Weber,~A.; others Oxide-based solid-state batteries: a perspective on composite cathode architecture. \emph{Advanced Energy Materials} \textbf{2023}, \emph{13}, 2201939\relax
\mciteBstWouldAddEndPuncttrue
\mciteSetBstMidEndSepPunct{\mcitedefaultmidpunct}
{\mcitedefaultendpunct}{\mcitedefaultseppunct}\relax
\EndOfBibitem
\bibitem[Minnmann \latin{et~al.}(2022)Minnmann, Strauss, Bielefeld, Ruess, Adelhelm, Burkhardt, Dreyer, Trevisanello, Ehrenberg, Brezesinski, \latin{et~al.} others]{minnmann2022designing}
Minnmann,~P.; Strauss,~F.; Bielefeld,~A.; Ruess,~R.; Adelhelm,~P.; Burkhardt,~S.; Dreyer,~S.~L.; Trevisanello,~E.; Ehrenberg,~H.; Brezesinski,~T.; others Designing cathodes and cathode active materials for solid-state batteries. \emph{Advanced Energy Materials} \textbf{2022}, \emph{12}, 2201425\relax
\mciteBstWouldAddEndPuncttrue
\mciteSetBstMidEndSepPunct{\mcitedefaultmidpunct}
{\mcitedefaultendpunct}{\mcitedefaultseppunct}\relax
\EndOfBibitem
\bibitem[Holzer \latin{et~al.}(2023)Holzer, Marmet, Fingerle, Wiegmann, Neumann, and Schmidt]{holzer2023tortuosity}
Holzer,~L.; Marmet,~P.; Fingerle,~M.; Wiegmann,~A.; Neumann,~M.; Schmidt,~V. Tortuosity and microstructure effects in porous media: classical theories, empirical data and modern methods. \emph{Springer} \textbf{2023}, \relax
\mciteBstWouldAddEndPunctfalse
\mciteSetBstMidEndSepPunct{\mcitedefaultmidpunct}
{}{\mcitedefaultseppunct}\relax
\EndOfBibitem
\bibitem[Minnmann \latin{et~al.}(2021)Minnmann, Quillman, Burkhardt, Richter, and Janek]{minnmann2021editors}
Minnmann,~P.; Quillman,~L.; Burkhardt,~S.; Richter,~F.~H.; Janek,~J. Editors’ choice—quantifying the impact of charge transport bottlenecks in composite cathodes of all-solid-state batteries. \emph{Journal of The Electrochemical Society} \textbf{2021}, \emph{168}, 040537\relax
\mciteBstWouldAddEndPuncttrue
\mciteSetBstMidEndSepPunct{\mcitedefaultmidpunct}
{\mcitedefaultendpunct}{\mcitedefaultseppunct}\relax
\EndOfBibitem
\bibitem[Finsterbusch \latin{et~al.}(2018)Finsterbusch, Danner, Tsai, Uhlenbruck, Latz, and Guillon]{finsterbusch2018high}
Finsterbusch,~M.; Danner,~T.; Tsai,~C.-L.; Uhlenbruck,~S.; Latz,~A.; Guillon,~O. High capacity garnet-based all-solid-state lithium batteries: fabrication and 3D-microstructure resolved modeling. \emph{ACS Applied Materials \& Interfaces} \textbf{2018}, \emph{10}, 22329--22339\relax
\mciteBstWouldAddEndPuncttrue
\mciteSetBstMidEndSepPunct{\mcitedefaultmidpunct}
{\mcitedefaultendpunct}{\mcitedefaultseppunct}\relax
\EndOfBibitem
\bibitem[Nam \latin{et~al.}(2018)Nam, Oh, Jung, and Jung]{nam2018toward}
Nam,~Y.~J.; Oh,~D.~Y.; Jung,~S.~H.; Jung,~Y.~S. Toward practical all-solid-state lithium-ion batteries with high energy density and safety: Comparative study for electrodes fabricated by dry-and slurry-mixing processes. \emph{Journal of Power Sources} \textbf{2018}, \emph{375}, 93--101\relax
\mciteBstWouldAddEndPuncttrue
\mciteSetBstMidEndSepPunct{\mcitedefaultmidpunct}
{\mcitedefaultendpunct}{\mcitedefaultseppunct}\relax
\EndOfBibitem
\bibitem[Zhang \latin{et~al.}(2017)Zhang, Weber, Weigand, Arlt, Manke, Schr\"oder, Koerver, Leichtweiss, Hartmann, Zeier, \latin{et~al.} others]{zhang2017interfacial}
Zhang,~W.; Weber,~D.~A.; Weigand,~H.; Arlt,~T.; Manke,~I.; Schr\"oder,~D.; Koerver,~R.; Leichtweiss,~T.; Hartmann,~P.; Zeier,~W.~G.; others Interfacial processes and influence of composite cathode microstructure controlling the performance of all-solid-state lithium batteries. \emph{ACS Applied Materials \& Interfaces} \textbf{2017}, \emph{9}, 17835--17845\relax
\mciteBstWouldAddEndPuncttrue
\mciteSetBstMidEndSepPunct{\mcitedefaultmidpunct}
{\mcitedefaultendpunct}{\mcitedefaultseppunct}\relax
\EndOfBibitem
\bibitem[Clausnitzer \latin{et~al.}(2023)Clausnitzer, M{\"u}cke, Al-Jaljouli, Hein, Finsterbusch, Danner, Fattakhova-Rohlfing, Guillon, and Latz]{clausnitzer2023optimizing}
Clausnitzer,~M.; M{\"u}cke,~R.; Al-Jaljouli,~F.; Hein,~S.; Finsterbusch,~M.; Danner,~T.; Fattakhova-Rohlfing,~D.; Guillon,~O.; Latz,~A. Optimizing the composite cathode microstructure in all-solid-state batteries by structure-resolved simulations. \emph{Batteries \& Supercaps} \textbf{2023}, e202300167\relax
\mciteBstWouldAddEndPuncttrue
\mciteSetBstMidEndSepPunct{\mcitedefaultmidpunct}
{\mcitedefaultendpunct}{\mcitedefaultseppunct}\relax
\EndOfBibitem
\bibitem[Neumann \latin{et~al.}(2021)Neumann, Hamann, Danner, Hein, Becker-Steinberger, Wachsman, and Latz]{neumann2021effect}
Neumann,~A.; Hamann,~T.~R.; Danner,~T.; Hein,~S.; Becker-Steinberger,~K.; Wachsman,~E.; Latz,~A. Effect of the 3D structure and grain boundaries on lithium transport in garnet solid electrolytes. \emph{ACS Applied Energy Materials} \textbf{2021}, \emph{4}, 4786--4804\relax
\mciteBstWouldAddEndPuncttrue
\mciteSetBstMidEndSepPunct{\mcitedefaultmidpunct}
{\mcitedefaultendpunct}{\mcitedefaultseppunct}\relax
\EndOfBibitem
\bibitem[Strauss \latin{et~al.}(2018)Strauss, Bartsch, de~Biasi, Kim, Janek, Hartmann, and Brezesinski]{strauss2018impact}
Strauss,~F.; Bartsch,~T.; de~Biasi,~L.; Kim,~A.-Y.; Janek,~J.; Hartmann,~P.; Brezesinski,~T. Impact of cathode material particle size on the capacity of bulk-type all-solid-state batteries. \emph{ACS Energy Letters} \textbf{2018}, \emph{3}, 992--996\relax
\mciteBstWouldAddEndPuncttrue
\mciteSetBstMidEndSepPunct{\mcitedefaultmidpunct}
{\mcitedefaultendpunct}{\mcitedefaultseppunct}\relax
\EndOfBibitem
\bibitem[Kaiser \latin{et~al.}(2018)Kaiser, Spannenberger, Schmitt, Cronau, Kato, and Roling]{kaiser2018ion}
Kaiser,~N.; Spannenberger,~S.; Schmitt,~M.; Cronau,~M.; Kato,~Y.; Roling,~B. Ion transport limitations in all-solid-state lithium battery electrodes containing a sulfide-based electrolyte. \emph{Journal of Power Sources} \textbf{2018}, \emph{396}, 175--181\relax
\mciteBstWouldAddEndPuncttrue
\mciteSetBstMidEndSepPunct{\mcitedefaultmidpunct}
{\mcitedefaultendpunct}{\mcitedefaultseppunct}\relax
\EndOfBibitem
\bibitem[Bielefeld \latin{et~al.}(2018)Bielefeld, Weber, and Janek]{bielefeld2018microstructural}
Bielefeld,~A.; Weber,~D.~A.; Janek,~J. Microstructural modeling of composite cathodes for all-solid-state batteries. \emph{The Journal of Physical Chemistry C} \textbf{2018}, \emph{123}, 1626--1634\relax
\mciteBstWouldAddEndPuncttrue
\mciteSetBstMidEndSepPunct{\mcitedefaultmidpunct}
{\mcitedefaultendpunct}{\mcitedefaultseppunct}\relax
\EndOfBibitem
\bibitem[Bielefeld \latin{et~al.}(2020)Bielefeld, Weber, and Janek]{bielefeld2020modeling}
Bielefeld,~A.; Weber,~D.~A.; Janek,~J. Modeling effective ionic conductivity and binder influence in composite cathodes for all-solid-state batteries. \emph{ACS Applied Materials \& Interfaces} \textbf{2020}, \emph{12}, 12821--12833\relax
\mciteBstWouldAddEndPuncttrue
\mciteSetBstMidEndSepPunct{\mcitedefaultmidpunct}
{\mcitedefaultendpunct}{\mcitedefaultseppunct}\relax
\EndOfBibitem
\bibitem[Bielefeld \latin{et~al.}(2022)Bielefeld, Weber, Rue{\ss}, Glavas, and Janek]{bielefeld2022influence}
Bielefeld,~A.; Weber,~D.~A.; Rue{\ss},~R.; Glavas,~V.; Janek,~J. Influence of lithium ion kinetics, particle morphology and voids on the electrochemical performance of composite cathodes for all-solid-state batteries. \emph{Journal of The Electrochemical Society} \textbf{2022}, \emph{169}, 020539\relax
\mciteBstWouldAddEndPuncttrue
\mciteSetBstMidEndSepPunct{\mcitedefaultmidpunct}
{\mcitedefaultendpunct}{\mcitedefaultseppunct}\relax
\EndOfBibitem
\bibitem[Froboese \latin{et~al.}(2019)Froboese, van~der Sichel, Loellhoeffel, Helmers, and Kwade]{froboese2019effect}
Froboese,~L.; van~der Sichel,~J.~F.; Loellhoeffel,~T.; Helmers,~L.; Kwade,~A. Effect of microstructure on the ionic conductivity of an all solid-state battery electrode. \emph{Journal of the Electrochemical Society} \textbf{2019}, \emph{166}, A318\relax
\mciteBstWouldAddEndPuncttrue
\mciteSetBstMidEndSepPunct{\mcitedefaultmidpunct}
{\mcitedefaultendpunct}{\mcitedefaultseppunct}\relax
\EndOfBibitem
\bibitem[Shi \latin{et~al.}(2020)Shi, Tu, Tian, Xiao, Miara, Kononova, and Ceder]{shi2020high}
Shi,~T.; Tu,~Q.; Tian,~Y.; Xiao,~Y.; Miara,~L.~J.; Kononova,~O.; Ceder,~G. High active material loading in all-solid-state battery electrode via particle size optimization. \emph{Advanced Energy Materials} \textbf{2020}, \emph{10}, 1902881\relax
\mciteBstWouldAddEndPuncttrue
\mciteSetBstMidEndSepPunct{\mcitedefaultmidpunct}
{\mcitedefaultendpunct}{\mcitedefaultseppunct}\relax
\EndOfBibitem
\bibitem[Cronau \latin{et~al.}(2022)Cronau, Duchardt, Szabo, and Roling]{cronau2022ionic}
Cronau,~M.; Duchardt,~M.; Szabo,~M.; Roling,~B. Ionic conductivity versus particle size of ball-milled sulfide-based solid electrolytes: strategy towards optimized composite cathode performance in all-solid-state batteries. \emph{Batteries \& Supercaps} \textbf{2022}, \emph{5}, e202200041\relax
\mciteBstWouldAddEndPuncttrue
\mciteSetBstMidEndSepPunct{\mcitedefaultmidpunct}
{\mcitedefaultendpunct}{\mcitedefaultseppunct}\relax
\EndOfBibitem
\bibitem[Yu and Siegel(2017)Yu, and Siegel]{yu2017grain}
Yu,~S.; Siegel,~D.~J. Grain boundary contributions to Li-ion transport in the solid electrolyte \(\mathrm{ Li_7La_3Zr_2O_{12}} \) (LLZO). \emph{Chemistry of Materials} \textbf{2017}, \emph{29}, 9639--9647\relax
\mciteBstWouldAddEndPuncttrue
\mciteSetBstMidEndSepPunct{\mcitedefaultmidpunct}
{\mcitedefaultendpunct}{\mcitedefaultseppunct}\relax
\EndOfBibitem
\bibitem[Ates \latin{et~al.}(2022)Ates, Neumann, Danner, Latz, Zarrabeitia, Stepien, Varzi, and Passerini]{ates2022elucidating}
Ates,~T.; Neumann,~A.; Danner,~T.; Latz,~A.; Zarrabeitia,~M.; Stepien,~D.; Varzi,~A.; Passerini,~S. Elucidating the role of microstructure in thiophosphate electrolytes--a combined experimental and theoretical study of $\beta$-\(\mathrm{Li_3PS_4}\). \emph{Advanced Science} \textbf{2022}, 2105234\relax
\mciteBstWouldAddEndPuncttrue
\mciteSetBstMidEndSepPunct{\mcitedefaultmidpunct}
{\mcitedefaultendpunct}{\mcitedefaultseppunct}\relax
\EndOfBibitem
\bibitem[Smyrek \latin{et~al.}(2015)Smyrek, Pr{\"o}ll, Seifert, and Pfleging]{smyrek2015laser}
Smyrek,~P.; Pr{\"o}ll,~J.; Seifert,~H.; Pfleging,~W. Laser-induced breakdown spectroscopy of laser-structured \(\mathrm{Li(NiMnCo)O_2}\) electrodes for lithium-ion batteries. \emph{Journal of the Electrochemical Society} \textbf{2015}, \emph{163}, A19\relax
\mciteBstWouldAddEndPuncttrue
\mciteSetBstMidEndSepPunct{\mcitedefaultmidpunct}
{\mcitedefaultendpunct}{\mcitedefaultseppunct}\relax
\EndOfBibitem
\bibitem[Mangang \latin{et~al.}(2016)Mangang, Seifert, and Pfleging]{mangang2016influence}
Mangang,~M.; Seifert,~H.; Pfleging,~W. Influence of laser pulse duration on the electrochemical performance of laser structured \(\mathrm{LiFePO_4}\) composite electrodes. \emph{Journal of Power Sources} \textbf{2016}, \emph{304}, 24--32\relax
\mciteBstWouldAddEndPuncttrue
\mciteSetBstMidEndSepPunct{\mcitedefaultmidpunct}
{\mcitedefaultendpunct}{\mcitedefaultseppunct}\relax
\EndOfBibitem
\bibitem[Kim \latin{et~al.}(2018)Kim, Drews, Chandrasekaran, Miller, and Sakamoto]{kim2018improving}
Kim,~Y.; Drews,~A.; Chandrasekaran,~R.; Miller,~T.; Sakamoto,~J. Improving Li-ion battery charge rate acceptance through highly ordered hierarchical electrode design. \emph{Ionics} \textbf{2018}, \emph{24}, 2935--2943\relax
\mciteBstWouldAddEndPuncttrue
\mciteSetBstMidEndSepPunct{\mcitedefaultmidpunct}
{\mcitedefaultendpunct}{\mcitedefaultseppunct}\relax
\EndOfBibitem
\bibitem[Habedank \latin{et~al.}(2018)Habedank, Kraft, Rheinfeld, Krezdorn, Jossen, and Zaeh]{habedank2018increasing}
Habedank,~J.~B.; Kraft,~L.; Rheinfeld,~A.; Krezdorn,~C.; Jossen,~A.; Zaeh,~M.~F. Increasing the discharge rate capability of lithium-ion cells with laser-structured graphite anodes: Modeling and simulation. \emph{Journal of The Electrochemical Society} \textbf{2018}, \emph{165}, A1563\relax
\mciteBstWouldAddEndPuncttrue
\mciteSetBstMidEndSepPunct{\mcitedefaultmidpunct}
{\mcitedefaultendpunct}{\mcitedefaultseppunct}\relax
\EndOfBibitem
\bibitem[Habedank \latin{et~al.}(2019)Habedank, Kriegler, and Zaeh]{habedank2019enhanced}
Habedank,~J.~B.; Kriegler,~J.; Zaeh,~M.~F. Enhanced fast charging and reduced lithium-plating by laser-structured anodes for lithium-ion batteries. \emph{Journal of The Electrochemical Society} \textbf{2019}, \emph{166}, A3940\relax
\mciteBstWouldAddEndPuncttrue
\mciteSetBstMidEndSepPunct{\mcitedefaultmidpunct}
{\mcitedefaultendpunct}{\mcitedefaultseppunct}\relax
\EndOfBibitem
\bibitem[Kraft \latin{et~al.}(2020)Kraft, Habedank, Frank, Rheinfeld, and Jossen]{kraft2020modeling}
Kraft,~L.; Habedank,~J.~B.; Frank,~A.; Rheinfeld,~A.; Jossen,~A. Modeling and simulation of pore morphology modifications using laser-structured graphite anodes in lithium-ion batteries. \emph{Journal of The Electrochemical Society} \textbf{2020}, \emph{167}, 013506\relax
\mciteBstWouldAddEndPuncttrue
\mciteSetBstMidEndSepPunct{\mcitedefaultmidpunct}
{\mcitedefaultendpunct}{\mcitedefaultseppunct}\relax
\EndOfBibitem
\bibitem[Chen \latin{et~al.}(2020)Chen, Namkoong, Goel, Yang, Kazemiabnavi, Mortuza, Kazyak, Mazumder, Thornton, Sakamoto, \latin{et~al.} others]{chen2020efficient}
Chen,~K.-H.; Namkoong,~M.~J.; Goel,~V.; Yang,~C.; Kazemiabnavi,~S.; Mortuza,~S.; Kazyak,~E.; Mazumder,~J.; Thornton,~K.; Sakamoto,~J.; others Efficient fast-charging of lithium-ion batteries enabled by laser-patterned three-dimensional graphite anode architectures. \emph{Journal of Power Sources} \textbf{2020}, \emph{471}, 228475\relax
\mciteBstWouldAddEndPuncttrue
\mciteSetBstMidEndSepPunct{\mcitedefaultmidpunct}
{\mcitedefaultendpunct}{\mcitedefaultseppunct}\relax
\EndOfBibitem
\bibitem[Kriegler \latin{et~al.}(2021)Kriegler, Hille, Stock, Kraft, Hagemeister, Habedank, Jossen, and Zaeh]{kriegler2021enhanced}
Kriegler,~J.; Hille,~L.; Stock,~S.; Kraft,~L.; Hagemeister,~J.; Habedank,~J.~B.; Jossen,~A.; Zaeh,~M.~F. Enhanced performance and lifetime of lithium-ion batteries by laser structuring of graphite anodes. \emph{Applied Energy} \textbf{2021}, \emph{303}, 117693\relax
\mciteBstWouldAddEndPuncttrue
\mciteSetBstMidEndSepPunct{\mcitedefaultmidpunct}
{\mcitedefaultendpunct}{\mcitedefaultseppunct}\relax
\EndOfBibitem
\bibitem[De~Lauri \latin{et~al.}(2021)De~Lauri, Krumbein, Hein, Prifling, Schmidt, Danner, and Latz]{de2021beneficial}
De~Lauri,~V.; Krumbein,~L.; Hein,~S.; Prifling,~B.; Schmidt,~V.; Danner,~T.; Latz,~A. Beneficial effects of three-dimensional structured electrodes for the fast charging of lithium-ion batteries. \emph{ACS Applied Energy Materials} \textbf{2021}, \emph{4}, 13847--13859\relax
\mciteBstWouldAddEndPuncttrue
\mciteSetBstMidEndSepPunct{\mcitedefaultmidpunct}
{\mcitedefaultendpunct}{\mcitedefaultseppunct}\relax
\EndOfBibitem
\bibitem[Goel \latin{et~al.}(2023)Goel, Chen, Dasgupta, and Thornton]{goel2023optimization}
Goel,~V.; Chen,~K.-H.; Dasgupta,~N.~P.; Thornton,~K. Optimization of laser-patterned electrode architectures for fast charging of Li-ion batteries using simulations parameterized by machine learning. \emph{Energy Storage Materials} \textbf{2023}, \emph{57}, 44--58\relax
\mciteBstWouldAddEndPuncttrue
\mciteSetBstMidEndSepPunct{\mcitedefaultmidpunct}
{\mcitedefaultendpunct}{\mcitedefaultseppunct}\relax
\EndOfBibitem
\bibitem[Gottschalk \latin{et~al.}(2023)Gottschalk, Oertel, Strzelczyk, M{\"u}ller, Kr{\"u}ger, Haselrieder, and Kwade]{gottschalk2023improving}
Gottschalk,~L.; Oertel,~C.; Strzelczyk,~N.; M{\"u}ller,~J.; Kr{\"u}ger,~J.; Haselrieder,~W.; Kwade,~A. Improving the performance of lithium-ion batteries using a two-layer, hard carbon-containing silicon anode for use in high-energy electrodes. \emph{Energy Technology} \textbf{2023}, \emph{11}, 2200858\relax
\mciteBstWouldAddEndPuncttrue
\mciteSetBstMidEndSepPunct{\mcitedefaultmidpunct}
{\mcitedefaultendpunct}{\mcitedefaultseppunct}\relax
\EndOfBibitem
\bibitem[Chen \latin{et~al.}(2016)Chen, Liu, Liu, Tiu, Yang, Chu, and Wan]{chen2016improvement}
Chen,~L.-C.; Liu,~D.; Liu,~T.-J.; Tiu,~C.; Yang,~C.-R.; Chu,~W.-B.; Wan,~C.-C. Improvement of lithium-ion battery performance using a two-layered cathode by simultaneous slot-die coating. \emph{Journal of Energy Storage} \textbf{2016}, \emph{5}, 156--162\relax
\mciteBstWouldAddEndPuncttrue
\mciteSetBstMidEndSepPunct{\mcitedefaultmidpunct}
{\mcitedefaultendpunct}{\mcitedefaultseppunct}\relax
\EndOfBibitem
\bibitem[Wood \latin{et~al.}(2021)Wood, Li, Du, Daniel, Dunlop, Polzin, Jansen, Krumdick, and Wood~III]{wood2021impact}
Wood,~M.; Li,~J.; Du,~Z.; Daniel,~C.; Dunlop,~A.~R.; Polzin,~B.~J.; Jansen,~A.~N.; Krumdick,~G.~K.; Wood~III,~D.~L. Impact of secondary particle size and two-layer architectures on the high-rate performance of thick electrodes in lithium-ion battery pouch cells. \emph{Journal of Power Sources} \textbf{2021}, \emph{515}, 230429\relax
\mciteBstWouldAddEndPuncttrue
\mciteSetBstMidEndSepPunct{\mcitedefaultmidpunct}
{\mcitedefaultendpunct}{\mcitedefaultseppunct}\relax
\EndOfBibitem
\bibitem[Cheng \latin{et~al.}(2020)Cheng, Drummond, Duncan, and Grant]{cheng2020combining}
Cheng,~C.; Drummond,~R.; Duncan,~S.~R.; Grant,~P.~S. Combining composition graded positive and negative electrodes for higher performance Li-ion batteries. \emph{Journal of Power Sources} \textbf{2020}, \emph{448}, 227376\relax
\mciteBstWouldAddEndPuncttrue
\mciteSetBstMidEndSepPunct{\mcitedefaultmidpunct}
{\mcitedefaultendpunct}{\mcitedefaultseppunct}\relax
\EndOfBibitem
\bibitem[Rosen \latin{et~al.}(2022)Rosen, Finsterbusch, Guillon, and Fattakhova-Rohlfing]{rosen2022free}
Rosen,~M.; Finsterbusch,~M.; Guillon,~O.; Fattakhova-Rohlfing,~D. Free standing dual phase cathode tapes--scalable fabrication and microstructure optimization of garnet-based ceramic cathodes. \emph{Journal of Materials Chemistry A} \textbf{2022}, \emph{10}, 2320--2326\relax
\mciteBstWouldAddEndPuncttrue
\mciteSetBstMidEndSepPunct{\mcitedefaultmidpunct}
{\mcitedefaultendpunct}{\mcitedefaultseppunct}\relax
\EndOfBibitem
\bibitem[Knorr \latin{et~al.}(2022)Knorr, Hein, Prifling, Neumann, Danner, Schmidt, and Latz]{knorr2022}
Knorr,~T.; Hein,~S.; Prifling,~B.; Neumann,~M.; Danner,~T.; Schmidt,~V.; Latz,~A. Simulation-based and data-driven techniques for quantifying the influence of the carbon binder domain on electrochemical properties of Li-ion batteries. \emph{Energies} \textbf{2022}, \emph{15}\relax
\mciteBstWouldAddEndPuncttrue
\mciteSetBstMidEndSepPunct{\mcitedefaultmidpunct}
{\mcitedefaultendpunct}{\mcitedefaultseppunct}\relax
\EndOfBibitem
\bibitem[Yamamoto \latin{et~al.}(2018)Yamamoto, Terauchi, Sakuda, and Takahashi]{yamamoto2018binder}
Yamamoto,~M.; Terauchi,~Y.; Sakuda,~A.; Takahashi,~M. Binder-free sheet-type all-solid-state batteries with enhanced rate capabilities and high energy densities. \emph{Scientific Reports} \textbf{2018}, \emph{8}, 1212\relax
\mciteBstWouldAddEndPuncttrue
\mciteSetBstMidEndSepPunct{\mcitedefaultmidpunct}
{\mcitedefaultendpunct}{\mcitedefaultseppunct}\relax
\EndOfBibitem
\bibitem[Hippauf \latin{et~al.}(2019)Hippauf, Schumm, Doerfler, Althues, Fujiki, Shiratsuchi, Tsujimura, Aihara, and Kaskel]{hippauf2019overcoming}
Hippauf,~F.; Schumm,~B.; Doerfler,~S.; Althues,~H.; Fujiki,~S.; Shiratsuchi,~T.; Tsujimura,~T.; Aihara,~Y.; Kaskel,~S. Overcoming binder limitations of sheet-type solid-state cathodes using a solvent-free dry-film approach. \emph{Energy Storage Materials} \textbf{2019}, \emph{21}, 390--398\relax
\mciteBstWouldAddEndPuncttrue
\mciteSetBstMidEndSepPunct{\mcitedefaultmidpunct}
{\mcitedefaultendpunct}{\mcitedefaultseppunct}\relax
\EndOfBibitem
\bibitem[bes()]{bestwebsite}
BEST – Battery and Electrochemistry Simulation Tool - Fraunhofer ITWM. \emph{https://www.itwm.fraunhofer.de/en/departments/sms/products-services/best-battery-electrochemistry-simulation-tool.html} \relax
\mciteBstWouldAddEndPunctfalse
\mciteSetBstMidEndSepPunct{\mcitedefaultmidpunct}
{}{\mcitedefaultseppunct}\relax
\EndOfBibitem
\bibitem[Latz and Zausch(2011)Latz, and Zausch]{latz2011thermodynamic}
Latz,~A.; Zausch,~J. Thermodynamic consistent transport theory of Li-ion batteries. \emph{Journal of Power Sources} \textbf{2011}, \emph{196}, 3296--3302\relax
\mciteBstWouldAddEndPuncttrue
\mciteSetBstMidEndSepPunct{\mcitedefaultmidpunct}
{\mcitedefaultendpunct}{\mcitedefaultseppunct}\relax
\EndOfBibitem
\bibitem[Latz and Zausch(2015)Latz, and Zausch]{latz2015multiscale}
Latz,~A.; Zausch,~J. Multiscale modeling of lithium ion batteries: thermal aspects. \emph{Beilstein Journal of Nanotechnology} \textbf{2015}, \emph{6}, 987--1007\relax
\mciteBstWouldAddEndPuncttrue
\mciteSetBstMidEndSepPunct{\mcitedefaultmidpunct}
{\mcitedefaultendpunct}{\mcitedefaultseppunct}\relax
\EndOfBibitem
\bibitem[Newman and Balsara(2021)Newman, and Balsara]{newman2021electrochemical}
Newman,~J.; Balsara,~N.~P. \emph{Electrochemical systems}; John Wiley \& Sons, 2021\relax
\mciteBstWouldAddEndPuncttrue
\mciteSetBstMidEndSepPunct{\mcitedefaultmidpunct}
{\mcitedefaultendpunct}{\mcitedefaultseppunct}\relax
\EndOfBibitem
\bibitem[Danner \latin{et~al.}(2016)Danner, Singh, Hein, Kaiser, Hahn, and Latz]{danner2016thick}
Danner,~T.; Singh,~M.; Hein,~S.; Kaiser,~J.; Hahn,~H.; Latz,~A. Thick electrodes for Li-ion batteries: A model based analysis. \emph{Journal of Power Sources} \textbf{2016}, \emph{334}, 191--201\relax
\mciteBstWouldAddEndPuncttrue
\mciteSetBstMidEndSepPunct{\mcitedefaultmidpunct}
{\mcitedefaultendpunct}{\mcitedefaultseppunct}\relax
\EndOfBibitem
\bibitem[Amin and Chiang(2016)Amin, and Chiang]{amin2016characterization}
Amin,~R.; Chiang,~Y.-M. Characterization of electronic and ionic transport in \(\mathrm{Li_{1-x}Ni_{0.33}Mn_{0.33}Co_{0. 33}O_2}\) (NMC333) and \(\mathrm{Li_{1-x}Ni_{0.50}Mn_{0. 20}Co_{0.30}O_2}\) (NMC523) as a function of Li content. \emph{Journal of The Electrochemical Society} \textbf{2016}, \emph{163}, A1512\relax
\mciteBstWouldAddEndPuncttrue
\mciteSetBstMidEndSepPunct{\mcitedefaultmidpunct}
{\mcitedefaultendpunct}{\mcitedefaultseppunct}\relax
\EndOfBibitem
\bibitem[Ruess \latin{et~al.}(2020)Ruess, Schweidler, Hemmelmann, Conforto, Bielefeld, Weber, Sann, Elm, and Janek]{ruess2020influence}
Ruess,~R.; Schweidler,~S.; Hemmelmann,~H.; Conforto,~G.; Bielefeld,~A.; Weber,~D.~A.; Sann,~J.; Elm,~M.~T.; Janek,~J. Influence of NCM particle cracking on kinetics of lithium-ion batteries with liquid or solid electrolyte. \emph{Journal of The Electrochemical Society} \textbf{2020}, \emph{167}, 100532\relax
\mciteBstWouldAddEndPuncttrue
\mciteSetBstMidEndSepPunct{\mcitedefaultmidpunct}
{\mcitedefaultendpunct}{\mcitedefaultseppunct}\relax
\EndOfBibitem
\bibitem[Ryu \latin{et~al.}(2018)Ryu, Park, Yoon, and Sun]{ryu2018capacity}
Ryu,~H.-H.; Park,~K.-J.; Yoon,~C.~S.; Sun,~Y.-K. Capacity fading of Ni-rich \(\mathrm{Li[Ni_xCo_yMn_{1-x-y}] O_2}\) (\(\mathrm{0.6 \leq x \leq 0.95}\)) cathodes for high-energy-density lithium-ion batteries: bulk or surface degradation? \emph{Chemistry of Materials} \textbf{2018}, \emph{30}, 1155--1163\relax
\mciteBstWouldAddEndPuncttrue
\mciteSetBstMidEndSepPunct{\mcitedefaultmidpunct}
{\mcitedefaultendpunct}{\mcitedefaultseppunct}\relax
\EndOfBibitem
\bibitem[Nyman \latin{et~al.}(2008)Nyman, Behm, and Lindbergh]{nyman2008electrochemical}
Nyman,~A.; Behm,~M.; Lindbergh,~G. Electrochemical characterisation and modelling of the mass transport phenomena in \(\mathrm{LiPF_6}\)-EC-EMC electrolyte. \emph{Electrochimica Acta} \textbf{2008}, \emph{53}, 6356--6365\relax
\mciteBstWouldAddEndPuncttrue
\mciteSetBstMidEndSepPunct{\mcitedefaultmidpunct}
{\mcitedefaultendpunct}{\mcitedefaultseppunct}\relax
\EndOfBibitem
\bibitem[Landesfeind and Gasteiger(2019)Landesfeind, and Gasteiger]{landesfeind2019temperature}
Landesfeind,~J.; Gasteiger,~H.~A. Temperature and concentration dependence of the ionic transport properties of lithium-ion battery electrolytes. \emph{Journal of The Electrochemical Society} \textbf{2019}, \emph{166}, A3079--A3097\relax
\mciteBstWouldAddEndPuncttrue
\mciteSetBstMidEndSepPunct{\mcitedefaultmidpunct}
{\mcitedefaultendpunct}{\mcitedefaultseppunct}\relax
\EndOfBibitem
\bibitem[Vierrath \latin{et~al.}(2015)Vierrath, Zielke, Moroni, Mondon, Wheeler, Zengerle, and Thiele]{vierrath2015morphology}
Vierrath,~S.; Zielke,~L.; Moroni,~R.; Mondon,~A.; Wheeler,~D.~R.; Zengerle,~R.; Thiele,~S. Morphology of nanoporous carbon-binder domains in Li-ion batteries — A FIB-SEM study. \emph{Electrochemistry Communications} \textbf{2015}, \emph{60}, 176--179\relax
\mciteBstWouldAddEndPuncttrue
\mciteSetBstMidEndSepPunct{\mcitedefaultmidpunct}
{\mcitedefaultendpunct}{\mcitedefaultseppunct}\relax
\EndOfBibitem
\bibitem[Zielke \latin{et~al.}(2015)Zielke, Hutzenlaub, Wheeler, Chao, Manke, Hilger, Paust, Zengerle, and Thiele]{zielke2015three}
Zielke,~L.; Hutzenlaub,~T.; Wheeler,~D.~R.; Chao,~C.-W.; Manke,~I.; Hilger,~A.; Paust,~N.; Zengerle,~R.; Thiele,~S. Three-phase multiscale modeling of a \(\mathrm{LiCoO_2}\) cathode: combining the advantages of FIB--SEM imaging and X-ray tomography. \emph{Advanced Energy Materials} \textbf{2015}, \emph{5}, 1401612\relax
\mciteBstWouldAddEndPuncttrue
\mciteSetBstMidEndSepPunct{\mcitedefaultmidpunct}
{\mcitedefaultendpunct}{\mcitedefaultseppunct}\relax
\EndOfBibitem
\bibitem[Singh \latin{et~al.}(2015)Singh, Kaiser, and Hahna]{singh2015thick}
Singh,~M.; Kaiser,~J.; Hahna,~H. Thick electrodes for high energy lithium ion batteries. \emph{Journal of The Electrochemical Society} \textbf{2015}, \emph{162}, A1196--A1201\relax
\mciteBstWouldAddEndPuncttrue
\mciteSetBstMidEndSepPunct{\mcitedefaultmidpunct}
{\mcitedefaultendpunct}{\mcitedefaultseppunct}\relax
\EndOfBibitem
\bibitem[Diederichsen \latin{et~al.}(2017)Diederichsen, McShane, and McCloskey]{diederichsen2017promising}
Diederichsen,~K.~M.; McShane,~E.~J.; McCloskey,~B.~D. Promising routes to a high \(\mathrm{Li^+ }\) transference number electrolyte for lithium ion batteries. \emph{ACS Energy Letters} \textbf{2017}, \emph{2}, 2563--2575\relax
\mciteBstWouldAddEndPuncttrue
\mciteSetBstMidEndSepPunct{\mcitedefaultmidpunct}
{\mcitedefaultendpunct}{\mcitedefaultseppunct}\relax
\EndOfBibitem
\bibitem[Logan and Dahn(2020)Logan, and Dahn]{logan2020electrolyte}
Logan,~E.; Dahn,~J. Electrolyte design for fast-charging Li-ion batteries. \emph{Trends in Chemistry} \textbf{2020}, \emph{2}, 354--366\relax
\mciteBstWouldAddEndPuncttrue
\mciteSetBstMidEndSepPunct{\mcitedefaultmidpunct}
{\mcitedefaultendpunct}{\mcitedefaultseppunct}\relax
\EndOfBibitem
\bibitem[Zhou \latin{et~al.}(2023)Zhou, Zhang, Xiang, and Liu]{zhou2023strategies}
Zhou,~P.; Zhang,~X.; Xiang,~Y.; Liu,~K. Strategies to enhance \(\mathrm{Li^+}\) transference number in liquid electrolytes for better lithium batteries. \emph{Nano Research} \textbf{2023}, \emph{16}, 8055--8071\relax
\mciteBstWouldAddEndPuncttrue
\mciteSetBstMidEndSepPunct{\mcitedefaultmidpunct}
{\mcitedefaultendpunct}{\mcitedefaultseppunct}\relax
\EndOfBibitem
\bibitem[Jung \latin{et~al.}(2014)Jung, Gwon, Hong, Park, Seo, Kim, Hyun, Yang, and Kang]{jung2014understanding}
Jung,~S.-K.; Gwon,~H.; Hong,~J.; Park,~K.-Y.; Seo,~D.-H.; Kim,~H.; Hyun,~J.; Yang,~W.; Kang,~K. Understanding the degradation mechanisms of \(\mathrm{LiNi_{0.5}Co_{0.2}Mn_{0.3}O_2}\) cathode material in lithium ion batteries. \emph{Advanced Energy Materials} \textbf{2014}, \emph{4}, 1300787\relax
\mciteBstWouldAddEndPuncttrue
\mciteSetBstMidEndSepPunct{\mcitedefaultmidpunct}
{\mcitedefaultendpunct}{\mcitedefaultseppunct}\relax
\EndOfBibitem
\bibitem[Nazarenus \latin{et~al.}(2021)Nazarenus, Sun, Exner, Kita, and Moos]{nazarenus2021powder}
Nazarenus,~T.; Sun,~Y.; Exner,~J.; Kita,~J.; Moos,~R. Powder aerosol deposition as a method to produce garnet-type solid ceramic electrolytes: a study on electrochemical film properties and industrial applications. \emph{Energy Technology} \textbf{2021}, \emph{9}, 2100211\relax
\mciteBstWouldAddEndPuncttrue
\mciteSetBstMidEndSepPunct{\mcitedefaultmidpunct}
{\mcitedefaultendpunct}{\mcitedefaultseppunct}\relax
\EndOfBibitem
\bibitem[Usiskin and Maier(2020)Usiskin, and Maier]{usiskin2020guidelines}
Usiskin,~R.; Maier,~J. Guidelines for optimizing the architecture of battery insertion electrodes with ohmic surface, coating, or electrolyte resistances. \emph{Journal of The Electrochemical Society} \textbf{2020}, \emph{167}, 080505\relax
\mciteBstWouldAddEndPuncttrue
\mciteSetBstMidEndSepPunct{\mcitedefaultmidpunct}
{\mcitedefaultendpunct}{\mcitedefaultseppunct}\relax
\EndOfBibitem
\bibitem[Xu \latin{et~al.}(2014)Xu, Wang, Ding, Chen, Nasybulin, Zhang, and Zhang]{xu2014lithium}
Xu,~W.; Wang,~J.; Ding,~F.; Chen,~X.; Nasybulin,~E.; Zhang,~Y.; Zhang,~J.-G. Lithium metal anodes for rechargeable batteries. \emph{Energy \& Environmental Science} \textbf{2014}, \emph{7}, 513--537\relax
\mciteBstWouldAddEndPuncttrue
\mciteSetBstMidEndSepPunct{\mcitedefaultmidpunct}
{\mcitedefaultendpunct}{\mcitedefaultseppunct}\relax
\EndOfBibitem
\bibitem[mat(2020)]{materialsproject}
Materials data on \(\mathrm{Li_6PS_5Cl}\) by Materials Project, DOI: 10.17188/1316731. 2020\relax
\mciteBstWouldAddEndPuncttrue
\mciteSetBstMidEndSepPunct{\mcitedefaultmidpunct}
{\mcitedefaultendpunct}{\mcitedefaultseppunct}\relax
\EndOfBibitem
\end{mcitethebibliography}


\providecommand{\latin}[1]{#1}
\makeatletter
\providecommand{\doi}
  {\begingroup\let\do\@makeother\dospecials
  \catcode`\{=1 \catcode`\}=2 \doi@aux}
\providecommand{\doi@aux}[1]{\endgroup\texttt{#1}}
\makeatother
\providecommand*\mcitethebibliography{\thebibliography}
\csname @ifundefined\endcsname{endmcitethebibliography}  {\let\endmcitethebibliography\endthebibliography}{}
\begin{mcitethebibliography}{10}
\providecommand*\natexlab[1]{#1}
\providecommand*\mciteSetBstSublistMode[1]{}
\providecommand*\mciteSetBstMaxWidthForm[2]{}
\providecommand*\mciteBstWouldAddEndPuncttrue
  {\def\EndOfBibitem{\unskip.}}
\providecommand*\mciteBstWouldAddEndPunctfalse
  {\let\EndOfBibitem\relax}
\providecommand*\mciteSetBstMidEndSepPunct[3]{}
\providecommand*\mciteSetBstSublistLabelBeginEnd[3]{}
\providecommand*\EndOfBibitem{}
\mciteSetBstSublistMode{f}
\mciteSetBstMaxWidthForm{subitem}{(\alph{mcitesubitemcount})}
\mciteSetBstSublistLabelBeginEnd
  {\mcitemaxwidthsubitemform\space}
  {\relax}
  {\relax}

\bibitem[Neumann \latin{et~al.}(2021)Neumann, Hamann, Danner, Hein, Becker-Steinberger, Wachsman, and Latz]{neumann2021effect}
Neumann,~A.; Hamann,~T.~R.; Danner,~T.; Hein,~S.; Becker-Steinberger,~K.; Wachsman,~E.; Latz,~A. Effect of the 3D structure and grain boundaries on lithium transport in garnet solid electrolytes. \emph{ACS Applied Energy Materials} \textbf{2021}, \emph{4}, 4786--4804\relax
\mciteBstWouldAddEndPuncttrue
\mciteSetBstMidEndSepPunct{\mcitedefaultmidpunct}
{\mcitedefaultendpunct}{\mcitedefaultseppunct}\relax
\EndOfBibitem
\bibitem[Finsterbusch \latin{et~al.}(2018)Finsterbusch, Danner, Tsai, Uhlenbruck, Latz, and Guillon]{finsterbusch2018high}
Finsterbusch,~M.; Danner,~T.; Tsai,~C.-L.; Uhlenbruck,~S.; Latz,~A.; Guillon,~O. High capacity garnet-based all-solid-state lithium batteries: fabrication and 3D-microstructure resolved modeling. \emph{ACS Applied Materials \& Interfaces} \textbf{2018}, \emph{10}, 22329--22339\relax
\mciteBstWouldAddEndPuncttrue
\mciteSetBstMidEndSepPunct{\mcitedefaultmidpunct}
{\mcitedefaultendpunct}{\mcitedefaultseppunct}\relax
\EndOfBibitem
\bibitem[Bielefeld \latin{et~al.}(2022)Bielefeld, Weber, Rue{\ss}, Glavas, and Janek]{bielefeld2022influence}
Bielefeld,~A.; Weber,~D.~A.; Rue{\ss},~R.; Glavas,~V.; Janek,~J. Influence of lithium ion kinetics, particle morphology and voids on the electrochemical performance of composite cathodes for all-solid-state batteries. \emph{Journal of The Electrochemical Society} \textbf{2022}, \emph{169}, 020539\relax
\mciteBstWouldAddEndPuncttrue
\mciteSetBstMidEndSepPunct{\mcitedefaultmidpunct}
{\mcitedefaultendpunct}{\mcitedefaultseppunct}\relax
\EndOfBibitem
\bibitem[Amin and Chiang(2016)Amin, and Chiang]{amin2016characterization}
Amin,~R.; Chiang,~Y.-M. Characterization of electronic and ionic transport in \(\mathrm{Li_{1-x}Ni_{0.33}Mn_{0.33}Co_{0. 33}O_2}\) (NMC333) and \(\mathrm{Li_{1-x}Ni_{0.50}Mn_{0. 20}Co_{0.30}O_2}\) (NMC523) as a function of Li content. \emph{Journal of The Electrochemical Society} \textbf{2016}, \emph{163}, A1512\relax
\mciteBstWouldAddEndPuncttrue
\mciteSetBstMidEndSepPunct{\mcitedefaultmidpunct}
{\mcitedefaultendpunct}{\mcitedefaultseppunct}\relax
\EndOfBibitem
\bibitem[Ruess \latin{et~al.}(2020)Ruess, Schweidler, Hemmelmann, Conforto, Bielefeld, Weber, Sann, Elm, and Janek]{ruess2020influence}
Ruess,~R.; Schweidler,~S.; Hemmelmann,~H.; Conforto,~G.; Bielefeld,~A.; Weber,~D.~A.; Sann,~J.; Elm,~M.~T.; Janek,~J. Influence of NCM particle cracking on kinetics of lithium-ion batteries with liquid or solid electrolyte. \emph{Journal of The Electrochemical Society} \textbf{2020}, \emph{167}, 100532\relax
\mciteBstWouldAddEndPuncttrue
\mciteSetBstMidEndSepPunct{\mcitedefaultmidpunct}
{\mcitedefaultendpunct}{\mcitedefaultseppunct}\relax
\EndOfBibitem
\bibitem[Nyman \latin{et~al.}(2008)Nyman, Behm, and Lindbergh]{nyman2008electrochemical}
Nyman,~A.; Behm,~M.; Lindbergh,~G. Electrochemical characterisation and modelling of the mass transport phenomena in \(\mathrm{LiPF_6}\)-EC-EMC electrolyte. \emph{Electrochimica Acta} \textbf{2008}, \emph{53}, 6356--6365\relax
\mciteBstWouldAddEndPuncttrue
\mciteSetBstMidEndSepPunct{\mcitedefaultmidpunct}
{\mcitedefaultendpunct}{\mcitedefaultseppunct}\relax
\EndOfBibitem
\bibitem[Landesfeind and Gasteiger(2019)Landesfeind, and Gasteiger]{landesfeind2019temperature}
Landesfeind,~J.; Gasteiger,~H.~A. Temperature and concentration dependence of the ionic transport properties of lithium-ion battery electrolytes. \emph{Journal of The Electrochemical Society} \textbf{2019}, \emph{166}, A3079--A3097\relax
\mciteBstWouldAddEndPuncttrue
\mciteSetBstMidEndSepPunct{\mcitedefaultmidpunct}
{\mcitedefaultendpunct}{\mcitedefaultseppunct}\relax
\EndOfBibitem
\bibitem[Xu \latin{et~al.}(2014)Xu, Wang, Ding, Chen, Nasybulin, Zhang, and Zhang]{xu2014lithium}
Xu,~W.; Wang,~J.; Ding,~F.; Chen,~X.; Nasybulin,~E.; Zhang,~Y.; Zhang,~J.-G. Lithium metal anodes for rechargeable batteries. \emph{Energy \& Environmental Science} \textbf{2014}, \emph{7}, 513--537\relax
\mciteBstWouldAddEndPuncttrue
\mciteSetBstMidEndSepPunct{\mcitedefaultmidpunct}
{\mcitedefaultendpunct}{\mcitedefaultseppunct}\relax
\EndOfBibitem
\bibitem[mat(2020)]{materialsproject}
Materials data on \(\mathrm{Li_6PS_5Cl}\) by Materials Project, DOI: 10.17188/1316731. 2020\relax
\mciteBstWouldAddEndPuncttrue
\mciteSetBstMidEndSepPunct{\mcitedefaultmidpunct}
{\mcitedefaultendpunct}{\mcitedefaultseppunct}\relax
\EndOfBibitem
\end{mcitethebibliography}

\end{document}


\clearpage

\subsection*{Simulation domain}

\begin{figure}[H]
    \centering
    \includegraphics[width=0.9\textwidth]{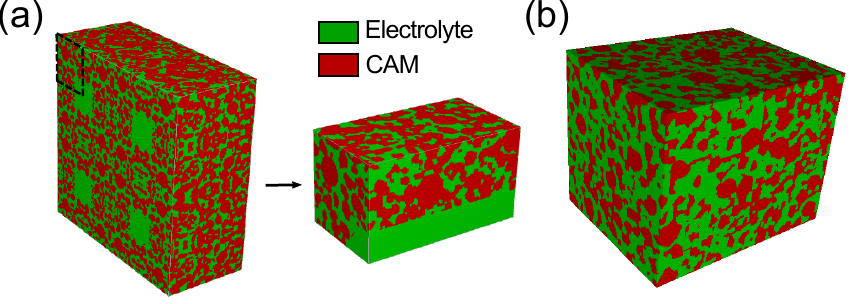}
    \caption{Overview of the generated 3D cathode structures. (a) Perforated cathode concept: To reduce computational costs, a quarter hole is simulated by taking advantage of the symmetry of the perforation pattern. (b) Layered cathode concept: A two-layered concept is investigated, with a first layer containing 60 vol\% CAM at the separator side and a second layer containing 70 vol\% CAM at the current-collector side.}
    \label{fig:3D_cathode_structure}
\end{figure}

\begin{figure}[H]
    \centering
    \includegraphics[width=0.7\textwidth]{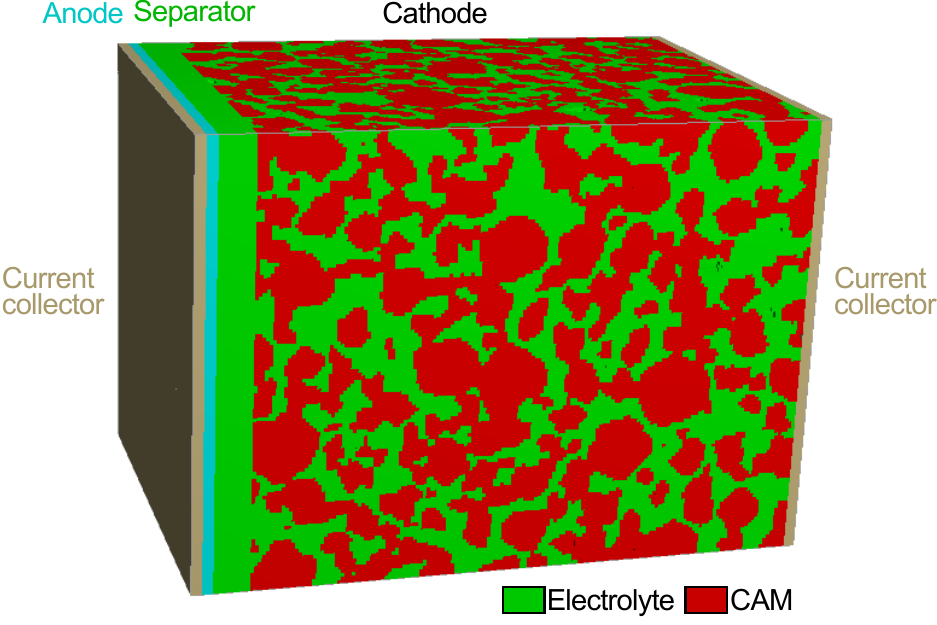}
    \caption{Overview of the input geometry for the structure-resolved simulations. A planar anode, separator, and current collectors are added to the generated cathode structures.}
    \label{fig:Overview_structure}
\end{figure}

\newpage

\subsection*{Parametrization}
\setcitestyle{numbers,open={[},close={]}}
\begin{table}
    \centering
    \begin{tabularx}{17cm}{ccccc}
\toprule
   Symbol & Value & Unit & Short description & Reference\\
   \midrule
   \textbf{Li-metal}\\
    $U_\text{0}^\text{An}$ & 0 & $\text{V}$ & Open circuit potential & -\\
    $\sigma_\text{Li}^\text{An}$ & $1$ & $S/cm$ & Electronic conductivity & -\\
    $i_0^\text{Li}$ & $2.59\cdot 10^{-2}$ & A/cm$^2$ & Exchange current density & \cite{neumann2021effect}\\
    $\alpha^\text{Li}$ & $0.5$ & - & Symmetry factor & \cite{finsterbusch2018high}\\
   \midrule
      \textbf{NMC 811}\\
    $U_{0}^\text{CAM}$ & $4.2$ & $\text{V}$ & Open circuit potential* & \cite{bielefeld2022influence}\\
    $c_\text{Li}^{CAM,0}$ & $0.01131$ & $mol/cm^3$ & Initial concentration of Li-ions & Calc.\\
    $c_\text{Li}^{CAM,max}$ & $0.04903$ & $mol/cm^3$ & Maximum concentration of Li-ions & Calc.\\ 
    $\sigma_\text{Li}^\text{CAM}$ & $8.83\cdot 10^{-3}$ & $S/cm$ & Electronic conductivity* & \cite{amin2016characterization}\\
    $D_\text{Li}^\text{CAM}$ & \makecell{LE: $1.63 \cdot 10^{-12}$ \\ SE: $8.71 \cdot 10^{-13}$} & $cm^2/s$ & Li-ion diffusion coefficient* & \cite{ruess2020influence}\\
    $i_{00}^\text{CAM}$ & \makecell{LE: $2.402\cdot 10^{-2}$ \\ SE: $1.5392 \cdot 10^{-3}$} & $\frac{\text{Acm}^{2.5}}{\text{mol}^{1.5}}$ & Exchange current density factor & Calc. from \cite{ruess2020influence}\\
    \midrule
   \textbf{LE} (\ce{LiPF_6})\\
   $c_\text{Li}^\text{LE}$ & $1\cdot10^{-3}$ & $mol/cm^3$ & Concentration of Li-ions & -\\
   $\kappa_\text{Li}^\text{LE}$ & $9.4 \cdot 10^{-3}$ & $S/cm$ & Li-ion bulk conductivity* & \cite{nyman2008electrochemical}\\
   $D_\text{Li}^\text{LE}$ & $3.79\cdot10^{-6}$ & $cm^2/s$ & Li-ion diffusion coefficient* & \cite{nyman2008electrochemical}\\
   $t_\text{Li}^{+}$ & $0.25$ & - & Transference number* & \cite{nyman2008electrochemical}\\
   TDF & $1.85$ & - & Thermodynamic factor* & \cite{landesfeind2019temperature}\\
    $l_\text{sep}$ & $20$ & $\mu \text{m}$ & Separator thickness & -\\
   \midrule
   \textbf{SE} (\ce{Li_6PS_5Cl})\\
   $c_\text{Li}^\text{SE}$ & $0.036662$ & $mol/cm^3$ & Concentration of Li-ions & Calc.\\
   $\kappa_\text{Li}^\text{SE}$ & $0.7\cdot 10^{-3}$ & $S/cm$ & Li-ion bulk conductivity & \cite{ruess2020influence}\\
   $t_\text{Li}^{+}$ & $1$ & - & Transference number & -\\
    $l_\text{sep}$ & $20$ & $\mu \text{m}$ & Separator thickness & -\\
   \midrule
   \textbf{Operation}\\
        $U_\text{cut}$ & $3.0$ & $\text{V}$ & Cut-off voltage & -\\
\bottomrule
\end{tabularx}
    \caption{Parameters of the electrochemical simulations. Functional parameters are indicated by * and are given at initial conditions.}
    \label{tab:Parameters-Simulation}
\end{table}

\setcitestyle{super,open={[},close={]}}

\newpage
\subsection{Performance indicators}

From our simulation results, we derive several performance indicators that are crucial for identifying limiting processes and evaluating cathode designs.

\paragraph{Effective ionic conductivity} The effective ionic conductivity of the cathode structures $\kappa_\text{eff}$ is determined by solving the Poisson equation for the electrolyte phase. We apply a voltage of $U=$1 V at the boundaries of the structure. From the resulting current density $i$ and the length of the cathode $l$, the effective ionic conductivity can be calculated using Equation \ref{equ:kappa_eff}.

\begin{equation}
    \kappa_\text{eff}=l\cdot\frac{i}{U}
    \label{equ:kappa_eff}
\end{equation}

\paragraph{Theoretical capacity} The theoretical capacity of the cathode structures is calculated based on their CAM fraction: 

\begin{equation}
C_\text{theo} = \frac{(c_{\text{max}} - c_\text{0}) \cdot F \cdot V_\text{CAM}}{A}
    \label{equ:C_theo}
\end{equation}

\paragraph{Energy density} An important performance indicator is the energy density of the battery cell, given by Equation \ref{equ:E_m}. The mass of the separator and cathode are deduced from the input microstructure used in our simulations. The anode mass is estimated, assuming an ideal matching between the negative and positive electrode. Please note that we neglect the weight of current collectors and cell housing. The additional parameters used to calculate the energy density are summarized in Table S2.

\begin{equation}
    E_\text{grav}=\frac{\int_{t_\text{0}}^{t_\text{end}} i \cdot U dt}{m_\text{An}+m_\text{Sep}+m_\text{Ca}}
    \label{equ:E_m}
\end{equation}

\paragraph{CAM utilization} The utilization of the CAM $\eta_\text{CAM}$ can be evaluated from the current Li-concentration $c_\text{CAM}$ through Equation \ref{equ:Util_CAM}:

\begin{equation}
    \eta_\text{CAM}=\frac{c_\text{CAM}-c_\text{CAM,0}}{c_\text{CAM,max}-c_\text{CAM,0}}
    \label{equ:Util_CAM}
\end{equation}

\setcitestyle{numbers,open={[},close={]}}
\begin{table}
    \centering
    \begin{tabularx}{14cm}{lllll}
\toprule
   Symbol & Value & Unit & Short description & Reference\\
   \midrule
   \textbf{Li-metal}\\
    $C_\text{grav,Li}^\text{theo}$ & 3861 & $mAh/g$ & Theoretical gravimetric capacity & \cite{xu2014lithium}\\
  \textbf{SE} (\ce{Li_6PS_5Cl})\\
    $\rho_\text{SE}$ & $1.64$ & $g/cm^3$ & Gravimetric density & \cite{materialsproject}\\
   \textbf{NMC 811}\\
    $\rho_\text{CAM}$ & $4.77$ & $g/cm^3$ & Gravimetric density & \cite{bielefeld2022influence}\\
\bottomrule
\end{tabularx}
    \caption{Parameters used for the calculation of gravimetric capacity and energy density.}
    \label{tab:SI_Parameters-E_m}
\end{table}
\setcitestyle{super,open={[},close={]}}

\newpage

\subsection*{Perforated cathodes}

\begin{figure}[H]
    \centering
    \includegraphics[width=1\textwidth]{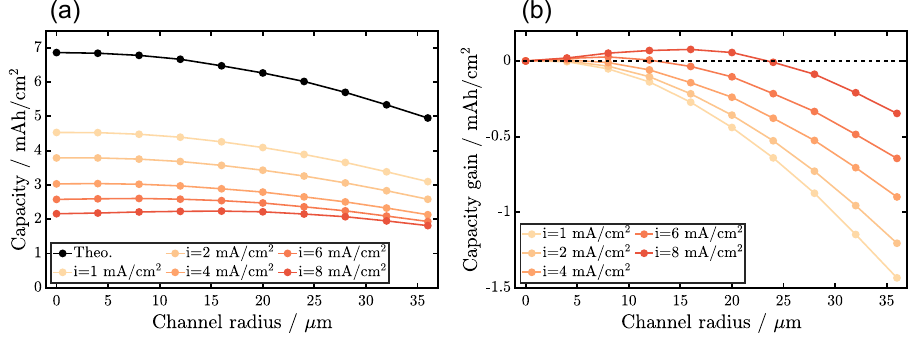}
    \caption{Effect of channel radius on capacity for the ASSB case. (a) Capacity for various current densities. The black line represents the theoretical capacity of the perforated structures. (b) Capacity gain of perforated structures at different current densities compared to the non-perforated structure ($r_\text{channel}=0$ $\mu m$). }
    \label{fig:SE_perf_cap}
\end{figure}

\newpage

\subsection*{Layered cathodes}

\begin{figure}[H]
    \centering
    \includegraphics[width=0.7\textwidth]{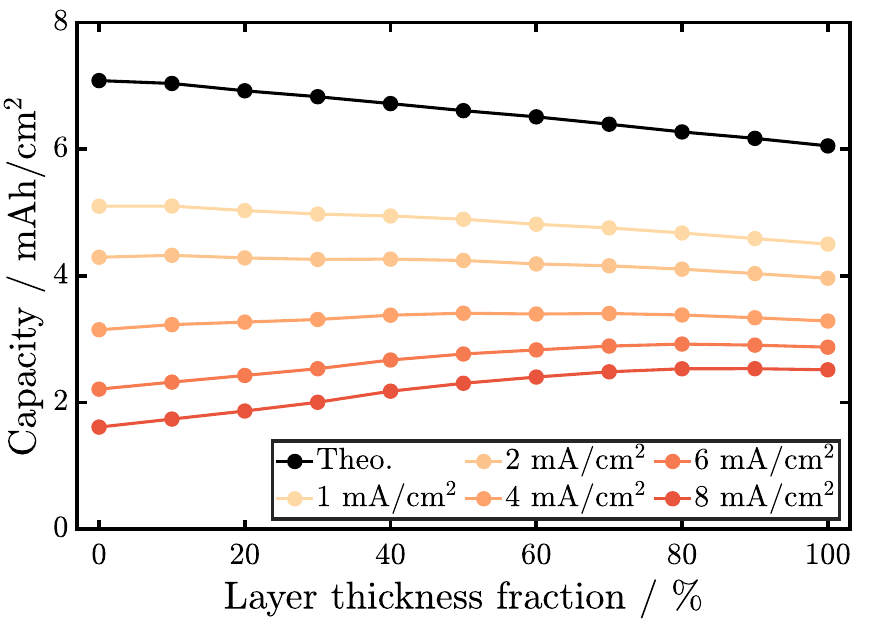}
    \caption{Influence of layer thickness fraction on ASSB capacity. Current densities range from 1 to 8 mA/cm². The black line represents the theoretical capacity of the generated structures.}
    \label{fig:SE_lay_cap}
\end{figure}

\begin{figure}[H]
    \centering
    \includegraphics[width=1\textwidth]{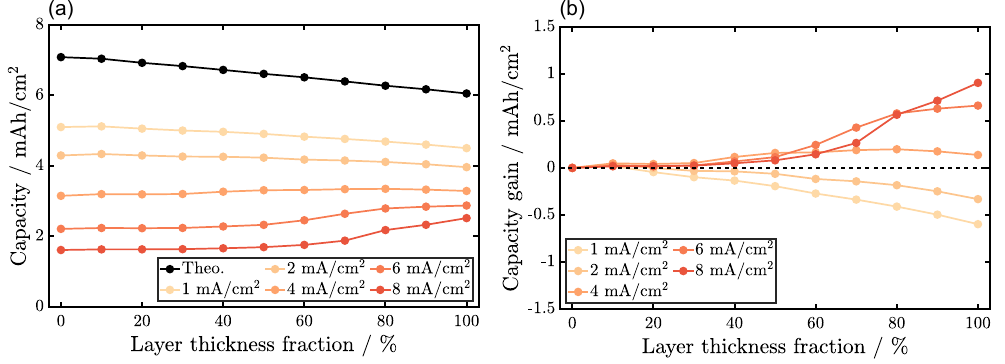}
    \caption{Influence of layer thickness fraction of the two-layered cathodes on electrochemical ASSB performance. The generated structures were reversed to show the significance of reducing tortuosity in the SE phase close to the separator. In the reversed structures, CAM loading is increased at the separator side and reduced at the current collector side of the cathode. (a) Capacity for current densities ranging from 1 to 8 mA/cm². The black line represents the theoretical capacity of the generated structures. (b) Capacity gain realized for the layered structures compared to a homogeneous cathode structure with 70 vol\% CAM ($f_\text{L60}=0$).}
    \label{fig:SE_lay_cap_turn}
\end{figure}
\newpage

\bibliography{literature.bib}